\documentstyle[aps,epsfig,floats]{revtex}

\tightenlines
\def\appendix{\par
 \setcounter{section}{0}
 \setcounter{subsection}{0}
 \def\thesection{Appendix \Alph{section}}
 \def\thesubsection{\Alph{section}.\arabic{subsection}}
 \def\theequation{\Alph{section}.\arabic{equation}}
 \setcounter{equation}{0}}
\newcommand{\be}{\begin{equation}}
\newcommand{\ee}{\end{equation}}
\newcommand{\bear}{\begin{eqnarray}}
\newcommand{\eear}{\end{eqnarray}}

\newcommand{\ip}{{\rm i}\epsilon}

\begin{document}
\draft
\title{Extraction of $\alpha_s$ from the Gross--Llewellyn Smith 
sum rule using Borel resummation}
\author{Carlos Contreras$^1$, Gorazd Cveti\v c$^1$, Kwang Sik Jeong$^2$ 
and  Taekoon Lee$^2$}
\address{$^1$Dept.~of Physics,
Universidad T\'ecnica Federico Santa Mar\'{\i}a,
Valpara\'{\i}so, Chile\\
$^2$Dept.~of Physics, Korea Advanced Institute of Science and Technology,
Daejon 305-701, Korea}
\maketitle
\begin{center}\today\end{center} 
\begin{abstract}
Using the CCFR data for the Gross--Llewellyn Smith (GLS) sum rule, 
we extract the strong coupling constant via Borel resummation of the 
perturbative QCD calculation. The method incorporates the correct 
nature of the first infrared renormalon singularity, and employs a 
conformal mapping to improve the convergence of the QCD perturbation 
expansion. The important twist--four contribution is {\it calculated} 
from resummation of the perturbation theory, which is based on the 
ansatz that the higher--twist contribution has a cut singularity only 
along the positive real axis on the complex coupling plane. Thus 
obtained, the strong coupling constant corresponding to the central 
GLS experimental value is in good agreement with the world average. 
\end{abstract}

\section{Introduction}
\label{introduction}

Many low--energy Quantum Chromodynamic (QCD) observables, including
the Gross--Llewellyn Smith (GLS) sum rule, with characteristic
energy scale a few GeVs are
analyzed in perturbative QCD within the framework of 
operator product expansion (OPE). 
In this scheme usually the most important is the perturbative
contribution from the Wilson coefficient 
of the unit operator, and there are
nonperturbative, power suppressed, higher--twist contributions.
Generally, the higher order coefficients in the
perturbative contribution grow rapidly due to the asymptotic nature
of the perturbative expansion. The uncalculated higher order
corrections are thus expected to be large, and this can cause a large 
uncertainty in data analysis that employs
the unprocessed, finite order perturbative expansion.
It is therefore important to properly handle the divergent
perturbation expansion via resummation, since it can give a more
accurate result with
reduced theoretical uncertainty. Besides, resummation serves to give a 
well defined meaning to the higher--twist contributions.
Without a proper resummation of the perturbative part, the 
higher--twist contributions are ambiguous 
\cite{david1,david2,mueller1,grunberg}.

An often used resummation technique is the Borel resummation.
It has a sound theoretical basis since it is built on our understanding
about the singularities in the Borel plane which cause the divergence
of the perturbative expansion.
Its use generally improves the quality of data analysis, as can be seen
from the reduced dependence on the renormalization scheme and scale, and
from the reduced dependence on the uncertainty of the uncalculated 
next higher order perturbative coefficient.
 
At moderate values of the strong coupling $\alpha_s(Q)$ at a few GeV,
the Borel integral receives most of its value from the interval
between the origin and the first infrared (IR) renormalon singularity,
and just beyond it, in the Borel plane. Let us  call this loosely defined
interval, for convenience,
the {\it primary interval}. 
In Borel resummation it is thus very important to describe the
Borel transform, which determines the Borel integral,
as accurately as possible in the primary interval
using the calculated first terms of perturbation theory.

For this purpose two steps can be taken: (1) an explicit incorporation
of the first IR renormalon singularity in the Borel transform, and (2)
use of an optimal conformal mapping.
With the usual power expansion of the Borel transform about the origin
the information on the renormalon singularity is lost. To remedy this,
one may explicitly incorporate the first renormalon singularity by writing
a Borel transform $\tilde D(b)$, which behaves as $1/(1-b/b_0)^{1+\nu}$
around the singularity at $b=b_0$, as
\be
\tilde D(b)= \frac{R(b)}{(1-b/b_0)^{1+\nu}}
\label{rb}
\ee
with $R(b)\equiv \tilde D(b) (1-b/b_0)^{1+\nu}$.
The function $R(b)$\footnote{This function was first introduced in
\cite{soper} in Borel resummation and independently in \cite{lee1}
in renormalon residue
calculation.}
is by definition bounded and has softer singularity at the
first IR renormalon.
Hence we can expect that 
the Borel transform in the form of (\ref{rb}),  with
$R(b)$ perturbatively expanded about the origin, would give a better
approximation than the direct expansion of $\tilde D(b)$.

The step (2) utilizes the information on the locations of the
singularities in the Borel plane. Use of conformal mapping in Borel
resummation has a long history \cite{zinn}, and its use 
in perturbative QCD was particularly
emphasized in \cite{mueller2}. On the Borel plane
there are IR renormalon singularities on the positive real axis
and ultraviolet (UV) renormalons on the negative axis.\footnote{
There are also instanton-caused singularities, which can be safely ignored
in our case.}
By pushing the singularities away from the primary interval one can
obtain a smoother $R$ in the new primary interval on the mapped plane.
This would render the perturbation of $R$ in the mapped plane 
to converge better.
Among the several mappings considered in the literature we find
that the one proposed in \cite{lee2} is particularly suited when used in
combination with
the Borel transform in the form of Eq.~(\ref{rb}). This mapping moves 
the first IR renormalon singularity that defines the primary interval 
to a point within the unit circle and all other singularities
to the unit circle.
Since the effect of the first IR renormalon is softened by the step (1),
we expect this mapping in our case
to be better suited than, for example, a mapping
that moves all singularities to the unit circle \cite{caprini}.

These techniques were applied to the hadronic tau decay width
\cite{lee2,caprini,lee3}
and to the hadronic contribution to 
the muon anomalous magnetic moment \cite{lee4}.
In this work we apply them to the GLS sum rule.
The CCFR analysis \cite{ccfr,jhkim} of the GLS sum rule was based on 
evaluation of the truncated perturbation series (TPS) in
${\overline {\rm MS}}$ scheme. Aside from the 
inherent ambiguity of the higher--twist 
contributions, this method gives predictions which
are not stable under the inclusion of an additional
term ($\sim \alpha_s^4$) in the TPS.
As we shall see, these problems can be avoided with
the use of Borel resummation.

A crucial new element of our analysis comes with the calculation
of the nonperturbative contribution.
Aside from the perturbative part, an important contribution to 
the GLS sum rule comes from the nonperturbative,
hadronic matrix element of the twist--four operator.
Being nonperturbative, this contribution is usually fitted using
the QCD sum rule calculation.
Recently it was proposed by one of us \cite{lee5},
motivated by an observation that  the 
nonperturbative amplitudes in lower dimensional solvable models have
a simple analyticity in complex coupling plane,
that these higher--twist contributions can
in principle be calculated from the Borel resummation of the perturbation
series. The proposal was based on the conjecture that
the higher--twist contributions have cuts only along the positive real
axis in the complex coupling plane, which allows to relate the real
part of the nonperturbative amplitude to its perturbatively
calculable imaginary part. This scheme was shown to work well in
model field theories. When applied to some of the
solvable lower dimensional theories, it allowed the associated
nonperturbative amplitudes to be accurately calculated from the first
terms of the perturbation theory in the respective theories.

{}From our analysis we obtain for the strong coupling parameter
the central value $\alpha_s(M_Z) \approx 0.117$.
Compared to the corresponding CCFR 
central value $\alpha_s(M_Z)=0.114$, 
our value is closer to the world average $\alpha_s(M_Z) \approx 0.118$.
The main improvement comes from  the correct incorporation of the
renormalon singularity on the Borel amplitude and the calculation of the
nonperturbative contribution.

The paper is organized in the following way.
In Sec.~\ref{method} we describe the resummation method,
incorporating in it the known structure of the
leading IR renormalon and the nonperturbative part,
as well as the conformal mapping. Section \ref{NA}
contains the numerical application of the method
to the GLS sum rule, leading to predictions for $\alpha_s(M_Z)$.
In Sec.~\ref{comparison} we compare our predictions
with those of other methods.
In Sec.~\ref{versus} we discuss some general features of the
Borel resummation and the OPE approaches to understand precisely
where the two methods deviate from each other.
In Sec.~\ref{conclusions}
we summarize our results and present conclusions.

\section{The method}
\label {method}

In this section we give an overview of our method used for the QCD analysis
of the GLS sum rule. Its implementation in detail will be given in the
following section. The QCD correction $\Delta (Q^2)$ to the GLS sum rule
is defined by
\be
\int_0^1 dx F_3^{\nu N}(x,Q^2)=3 (1-\Delta (Q^2))\,,
\ee
where $F_3^{\nu N}$ is the non-singlet deep inelastic scattering (DIS)
structure function in $\nu N$ scattering.
Here we shall ignore the target mass correction since it is irrelevant for 
our present discussion, but it will be included in the numerical analysis
presented in the next section.

We first begin by reviewing the old but important problem 
\cite{david1,david2} 
with  the conventional QCD formulation of  
$\Delta (Q^2)$ in OPE framework, which is 
widely used in data analysis. This problem is not confined to the
GLS sum rule, but generic to any perturbative OPE formulations.
$\Delta (Q^2)$ in OPE up to twist--four operator is given by
\be
\Delta (Q^2)= W_0(\alpha_s(Q)) +W_1(\alpha_s(Q)) \frac{\ll O \gg}{Q^2}\,,
\label{ope}
\ee
where $\alpha_s(Q)$ is the strong coupling constant, and 
$\ll \!\!O\!\!\gg$ is the reduced nucleonic matrix element 
of the twist--four operator that was first
derived in Ref.~\cite{Shuryak:1981kj}
\begin{eqnarray}
O_{\mu} & = & \bar{u} \tilde G_{\mu\nu} \gamma^{\nu} \gamma_5 u +
\bar{d} \tilde G_{\mu\nu} \gamma^{\nu}\gamma_5 d \ ,
\nonumber\\
{\tilde G}^{\mu\nu} &=& \frac{1}{2} \epsilon^{\mu \nu \alpha \beta}
G^{a}_{\alpha \beta} \frac{\lambda^a}{2} \ , \quad
\langle P | O_{\mu} | P \rangle_{\rm spin \ averaged} = 
2\; p_{\mu} \ll O \gg \ .
\nonumber
\end{eqnarray}
Here, $\lambda^{a}$ are the usual Gell-Mann matrices
and we used the notations of Ref.~\cite{braun-ko}.
Throughout the article we shall consider only the twist--four contribution
as the nonperturbative effect, and ignore higher twist contributions
since they are believed to be small.

In conventional QCD analysis the Wilson coefficients $W_i, (i=0,1)$, in
(\ref{ope}) are taken from the finite order, perturbative QCD calculation
in an usual renormalization scheme, say $\overline{\rm MS}$ ~scheme, and
the reduced matrix element $\ll \!\!O\!\!\gg$ from  
data fitting or QCD sum rule calculations, etc.
However, this scheme is, in principle, fundamentally flawed,
since perturbatively the
Wilson coefficients are not well--defined.
In perturbative calculation of the Wilson coefficients the quantum
fluctuations of all energy scale contribute, and in particular at large
orders the contribution from the far infrared regimes, where perturbative
QCD should fail, is large and gives rise to a same sign, factorially growing
large order behavior \cite{lautrup}.
Thus, without some kind of resummation of the divergent perturbation series,
the Wilson coefficients are not well--defined, and therefore
neither is the OPE
(\ref{ope}). As a consequence, no well--defined meaning that is 
independent of the 
definition of the Wilson coefficient $W_0$,
can be assigned to the matrix element $\ll \!\!O\!\!\gg$
\cite{david1,david2,mueller1}.

One could in principle introduce an infrared cutoff $\mu^2 (\ll Q^2)$ in the
perturbative calculation of $W_0$, and
regard the twist--four term to contain all the low momentum
$-k^2\!\equiv\!K^2 < \mu^2$ contributions \cite{nsvz,bb}.
This would remove the infrared renormalon and make the perturbation
expansion for $W_0$ convergent, and the twist--four term to be
$\mu^2$ dependent \cite{bb}.
However, to do this in practice is
impossible, because it is impossible to compute arbitrarily complicated
Feynman amplitudes of arbitrarily high power, and in particular
their small momentum contributions, as stressed by the authors of
Ref.~\cite{bb}. 

The problem discussed so far is not of academic nature only, 
but has important practical implications.
One might still think the OPE (\ref{ope}), with finite order perturbative
Wilson coefficients, is a good approximation scheme, since at any rate
the perturbative Wilson coefficients 
can be regarded as a good approximation at reasonably small
values of the strong coupling constant. Actually, this would be the case,
provided that the nonperturbative, higher--twist effect is far larger 
than the ignored next higher perturbative 
term in $W_0$. In practice, however, 
this condition is not supported by data
analysis. For instance, if this were the case, we would expect little
variation in the fitted values for the twist--four contribution over the
order of perturbation in $W_0$. But the variation is not small at all.
In the QCD sum rule calculation  using the
next--to--leading order (NLO) $W_0$ \cite{braun-ko},
the twist--four contribution was found 
not to be small, roughly equal to the
perturbative correction at $Q=1\,{\rm GeV}$. But, it was observed
in Refs.~\cite{KPS,yang} that 
the twist--four contribution virtually vanishes
when fitted with the next--to--next--to--leading order (NNLO) 
$W_0$ against the 
parton distribution functions extracted from experiments, 
and most of the twist--four contribution
extracted in the NLO fit can be accounted for by the perturbative 
NNLO contributions. This can be interpreted as a clear 
manifestation of the inherent ambiguity of the OPE approach.
Moreover, this tendency of strong dependence of higher--twist
contribution on the order of perturbation appears not to be special
to the GLS sum rule, but generic. The recent new estimate of the gluon
condensate \cite{ioffe},
from fitting the vectorial spectral function of hadronic
tau decay using NNLO Adler function, gives a small central value of
only $1/3$ that of the original QCD sum rule estimate \cite{svz} which uses
the leading order perturbation. These examples strongly indicate
that the inherent ambiguity of the OPE  can have
important consequences in practical applications, and demands
a careful treatment.

Borel resummation resolves this problem, 
which proceeds as follows \cite{zinn}.
The perturbation series for $W_0$
\be
W_0 = \sum_0^\infty w_n a(Q)^{n+1} \quad 
\left[ a(Q) \equiv \alpha_s(Q)/\pi \right] \ ,
\label{Dexp}
\ee
which is, being of same sign at large orders\footnote{Here we ignore,
for the moment, the UV renormalons, which give rise to sign-alternating 
large order behaviors. Being Borel summable, the UV renormalons can
in principle be treated separately, and do not affect our discussion 
in any essential way.}, 
non-Borel resummable at physical,
positive coupling $a(Q)$. 
So it is first Borel resummed at negative $a(Q)$
which yields a Borel resummed amplitude $\Delta_{\rm P}(a(Q))$. 
Then to obtain a Borel resummed physical amplitude one may analytically
continue
 $\Delta_{\rm P}$ to positive $a(Q)$ in the complex coupling plane.
This gives for $a(Q) >0$ \cite{lee5}
 \be
\Delta_{\rm P}(a(Q)\pm\ip)=\frac{1}{\beta_0} \int _{0\pm\ip}^{\infty\pm\ip}
db\, e^{-b/\beta_0 a(Q)}\,\tilde{W}_0(b)
\label{borelintegral}
\ee
with
\be
\tilde{W}_0(b)= \sum_{n=0}^\infty \frac{w_n}{n!}\, 
(b/\beta_0)^n \,.
\label{boreltransform}
\ee
Inserted for normalization convenience, $\beta_0$ is the one--loop
coefficient of the QCD $\beta$-function. Eq.~(\ref{borelintegral})
shows explicitly that in case of existing singularities on the
line of integration, i.e., IR renormalons, the real part of the
Borel integral $\Delta_{\rm P}$ is the (generalization of the)
Cauchy Principal Value. 
The Borel transform (\ref{boreltransform}) 
is believed to be convergent on
the unit disk $|b| < 1$, and is known to have a branch cut along
the positive real axis beginning at $b=1$. Near the branch cut 
it behaves as \cite{mueller1}
\be
\tilde{W}_0(b) = \frac{C}{\Gamma(-\nu)} \beta_0^{1+\nu}
\,(1-b)^{-1-\nu} (1+ O(1-b)) + \left(\text{Analytic 
part}\right)
\label{singularity}
\ee
with
\be
\nu = (\beta_1/\beta_0 -\gamma_2)/\beta_0\,,
\label{nu}
\ee
where $\beta_1$ and $\gamma_2$ are respectively
the two--loop coefficient of the QCD $\beta$--function and the
one--loop coefficient of the anomalous dimension of the 
twist--four operator appearing in the OPE (\ref{ope}). 
In our case $\nu$ is positive; for instance,  $\nu=32/81$ when three
quark flavors ($n_f\!=\!3$) are active \cite{Shuryak:1981kj}.\footnote{
According to Ref.~\cite{Shuryak:1981kj}:
$\gamma_2 = (N_c\!-\!1/N_c)/3 = 8/9$. Our convention for
parameters $\beta_j$ (and $c_j\!\equiv\!\beta_j/\beta_0$) 
is specified by Eq.~(\ref{RGE}) in the next Section.
For $n_f\!=\!3$, we have $\beta_0\!=\!9/4$, $\beta_1\!=\!4$.
Therefore, $\nu\!=\!32/81$.}  
Note that the Borel transform beyond the convergence disk of the series
(\ref{boreltransform}) can be obtained by analytic
continuation.
The analytic part,
which is analytic around the singularity, is not calculable, but the
residue $C$, a real number, is calculable perturbatively \cite{lee1,lee6}.

Because of the branch cut the Borel integrals in (\ref{borelintegral})
develop imaginary parts beginning at $b=1$.
Since the QCD correction $\Delta(Q^2)$ must be real, clearly
$\Delta_{\rm P}$ alone cannot reproduce the true amplitude.
There must be something else. Precisely at this point, the nonperturbative
amplitude, denoted $\Delta_{\rm NP}(a(Q))$,
comes to the rescue, which cancels the imaginary parts of $\Delta_{\rm P}$,
rendering the sum of the two to be real.
This $\Delta_{\rm NP}$ can be shown to be directly related to the twist--four
contribution in the OPE (\ref{ope}) \cite{parisi}, with its overall form
governed by the associated RG equation,
and may be regarded as a refinement of the latter, now free from the
inherent ambiguity of the OPE. 
The QCD correction may  now be written as
\bear
\Delta (Q^2)&=& \Delta_{\rm P} (a\pm\ip) +
\Delta_{\rm NP} (a\pm\ip) \nonumber \\
            &=& {\rm Re} [\Delta_{\rm P} (a\pm\ip)] +{\rm Re}
            [\Delta_{\rm NP}(a\pm\ip)] \,.
\label{formula}
\eear
Either sign can be taken because the result 
is independent of the sign chosen.

What can we tell about the nonperturbative amplitude $\Delta_{\rm NP}$?
Its imaginary part at a positive coupling is certainly calculable
from the perturbation theory because it is essentially the
imaginary part of $\Delta_{\rm P}$, albeit with opposite sign,
which is calculable in principle from the perturbation theory.
However, as we see in (\ref{formula}), what we need  is
the real part, which is certainly not directly calculable.

There is, however, an intriguing possibility of perturbative calculation
of the real part. From the perspective of Borel resummation 
the sole reason for the introduction of the 
nonperturbative amplitude above was to cancel the imaginary
parts arising from the analytic
continuation of  $\Delta_{\rm P}$ to physical coupling in the complex
coupling plane.
With the Borel integral (\ref{borelintegral})
and the singular Borel transform (\ref{singularity}), we can see 
that $\Delta_{\rm P}(a(Q))$, which by definition can have a singularity
only along the positive real axis in the complex $a(Q)$ plane,
has a branch cut  of the form (see \ref{appendix2})
\be
-C (-a(Q))^{-\nu} \,e^{-1/\beta_0 a(Q)}\, [\,1+ O(a)\,]\,.
\label{branchcut}
\ee
The nonperturbative amplitude $\Delta_{\rm NP}$ should cancel 
the imaginary part arising from this at positive coupling. 
The simplest functional form for $\Delta_{\rm NP}$ that can achieve
this purpose can be
obtained by postulating that, as has $\Delta_{\rm P}$, the nonperturbative
amplitude have a branch cut only along the positive real axis in the
coupling plane. This then leads to the following conjecture 
for $\Delta_{\rm NP}$ \cite{lee5}:
\be
\Delta_{\rm NP}(a(Q))=
C (-a(Q))^{-\nu} \,e^{-1/\beta_0 a(Q)}\, [\,1+ O(a)\,]
\label{np}
\ee
This has a very important implication because it allows to relate the
real part of the nonperturbative amplitude to the calculable imaginary
part. In this paper we  will 
adopt this conjecture, which appears very plausible,
at least to us, and as will be mentioned shortly it is supported by
some lower dimensional solvable models.
Now with (\ref{formula}) and (\ref{np}) 
we can write the QCD correction with only
the calculable $\Delta_{\rm P}$ as
\be
\Delta_{\rm P}(Q^2) + \Delta_{\rm NP}(Q^2) =
\left\{{\rm Re} \mp \cot(\nu\pi) \,{\rm Im}\right\} 
[\Delta_{\rm P} (a(Q)\pm\ip)] \,.
\label{masterequation}
\ee
This equation is the basis of our numerical analysis 
in the following section.
In \ref{appendix2} we present,
for reference, some explicit formulas
leading to (\ref{branchcut})--(\ref{masterequation}).

The argument that led to the determination of the nonperturbative
amplitude above did not rely on any special property of OPE, but
only on the general property of Borel resummation of a same sign
divergent perturbation theory. Thus it can be applied to any
perturbation theory with a same sign large order behavior.
Application of this scheme to the lower dimensional, solvable models
such as the double well potential and the two-dimensional 
nonlinear $\sigma$ model in large $N$ limit, allowed  an accurate 
calculation of the associated nonperturbative amplitude using only
the first terms of the perturbation expansion in the respective models
\cite{lee5}. Moreover, interestingly,
the numerical estimate of the gluon 
condensate from this scheme applied to the
Adler function gives a value virtually identical to the new estimate 
\cite{ioffe}.

Given the QCD correction  in the form (\ref{masterequation}) with
$\Delta_{\rm P} (a\pm\ip)$ given by (\ref{borelintegral}),
our aim is now to describe the Borel transform $\tilde W_0$ 
as accurately as possible in the primary interval 
using the known
first terms of the perturbation series.
For this, following the outline in Introduction, we write the Borel integral
(\ref{borelintegral}) as
 \be
\Delta_{\rm P}(a(Q)\pm\ip)=
\frac{1}{\beta_0} \int _{0\pm\ip}^{\infty\pm\ip}
db\, e^{-b/\beta_0 a(Q)}\,\frac{R(b)}{(1-b)^{1+\nu}}
\label{borelintegral1}
\ee
with $R(b)$ now defined by
\be
R(b)\equiv(1-b)^{1+\nu} \,\tilde W_0(b)
\label{rb-new}
\ee
and $\nu$ given in (\ref{nu}) ($\nu = 32/81$).
This step is expected to greatly improve the perturbative
description of the Borel transform in the primary interval because it
implements the renormalon singularity correctly, and renders us to deal
with a much softer singularity.
The singularity of $R(b)$ at $b=1$ is a branch cut and thus softer than
that of $\tilde W_0(b)$.

We can obtain further improvement by use of a conformal mapping that
exploits the known locations of the singularities of $\tilde W_0(b)$.
The latter is known to have renormalon singularities at non-zero 
integer values of $b$ on the real axis \cite{TM,Beneke:1999ui}. 
To speed up the convergence of the perturbative expansion of  $R(b)$, 
we may push the singularities, save the unavoidable first IR renormalon,
as far away from the origin as possible. This way we can reduce the
influence of the renormalon singularities and make $R(b(w))$ smoother 
around the primary interval.
One such a mapping we consider is \cite{lee2}
\be
w(b) = \frac{ \sqrt{1+b} - \sqrt{1 - b/2}}{ \sqrt{1+b} + \sqrt{1 - b/2}}
 \ \Rightarrow \ 
b(w) = \frac{8 w}{(3 w^2 - 2 w + 3)} \ .
\label{ct}
\end{equation}
This maps the first IR renormalon to $w=1/3$ and all other singularities
to the unit circle $|w|=1$.
We expect this mapping combined with the implementation (\ref{rb-new})
to provide an optimized environment for the Borel integral.
In the mapped plane the Borel integral now assumes the form
\be
\Delta_{\rm P}(a(Q)\pm\ip)=\frac{1}{\beta_0} \int _{{\cal C_\pm}}
dw \, e^{-b(w)/\beta_0 a(Q)}\, \frac{db(w)}{dw}\,
\frac{R(b(w))}{(1-b(w))^{1+\nu}} \,,
\label{borelintegral2}
\ee
where one of the integration contours ${\cal C_{-}}$ is shown in Fig. 1.
Again, since the answer is independent of the sign chosen, either contour
can be taken.
In the next section we perform a numerical analysis of
our implementation of the QCD correction: Eqs. (\ref{masterequation})
and (\ref{borelintegral2}).

\section{Numerical analysis}
\label{NA}

In this Section we will apply the method described in the previous 
Section to the Gross--Llewellyn Smith (GLS) sum rule, 
deducing values of the QCD coupling parameter 
$\alpha_s^{{\overline {\rm MS}}}(M_Z^2)$ from the GLS
values extracted from experiments.
In the first Subsection, we will present the resummed expression 
for the contributions of the three massless quark flavors.
As a byproduct of the obtained expression, we will obtain
an estimate of the next--to--next--to--next--to--leading 
(${\rm N}^3{\rm L}$) coefficient $w_3$ of the perturbative 
expansion. In the subsequent Subsection, we will include the
effects of the massive fourth quark flavor ($c$--quark)
to the GLS observable. In the last Subsection, the
available measured GLS values will be confronted with
our resummed expression and values of
$\alpha_s^{{\overline {\rm MS}}}(M_Z^2)$ will be
extracted.

\subsection{GLS -- the massless \lowercase{$n_f=3$} part} 
\label{massless}

The GLS quantity $M_3(Q^2)$ is the following integral (first moment)
of the charged--current non--polarized DIS structure function
$F_3\!\equiv\!(F_3^{\nu p} + F_3^{\nu n})/2$ 
over the Bjorken parameter $x$:
\begin{equation}
M_3(Q^2) = \frac{1}{3} \int_0^1 dx \; F_{3}(x;Q^2) 
\frac{\zeta^2}{x^2}
\left[ 1 + 2 \left( 1 + \frac{4 m_N^2 x^2}{Q^2} \right)^{1/2}
\right] \ ,
\label{M3o1}
\end{equation}
where $\zeta \equiv 2 x/(1 + \sqrt{1 + 4 m_N^2 x^2/Q^2})$
is the Nachtmann variable \cite{nachtmann}, and
$m_N$ is the nucleon mass. The quantity $M_3$
is the first Nachtmann moment of $F_3$ which absorbs all the
kinematical power corrections (target mass corrections: TMC)
$\sim (m_N^2/Q^2)^n$. The quantity $Q^2=-q^2$ is the
virtuality of the exchanged gauge boson; $Q^2$ 
characterizes the typical process momenta of the
observable $M_3(Q^2)$.
When expanding the integrand in
Eq.~(\ref{M3o1}) in powers of $m_N^2/Q^2$, we can rewrite
\begin{equation}
M_3(Q^2) =  \int_0^1 dx \; F_{3}(x;Q^2)
\left[ 1 - \frac{2}{3} \frac{m_N^2 x^2}{Q^2} +
\left( \frac{m_N^2 x^2}{Q^2} \right)^2 + {\cal O}\left(
\frac{m_N^6}{Q^6} \right) \right] \ .
\label{M3o2}
\end{equation}
The above GLS moment can be written in the form
\begin{equation}
M_3(Q^2) \equiv  3 \left( 1 - \Delta(Q^2) \right) \ ,
\label{M3o3}
\end{equation}
where the ``canonical'' quantity $\Delta(Q^2)$
has the following power expansion $W_0$
[cf.~Eq.~(\ref{ope})]:
\begin{equation}
\Delta(Q^2) \mapsto 
W_0(Q^2) = a (1 + w_1 a + w_2 a^2 + w_3 a^3 + \cdots ) \ .
\label{cGLSP1}
\end{equation}
Here, $a \equiv \alpha_s(\mu^2;c_2,c_3,\ldots)/\pi$
is the QCD coupling parameter with a given choice
of the renormalization scale (RScl) $\mu^2$ and
the renormalization scheme (RSch) parameters $c_j$ ($j \geq 2$).
The evolution of $a$ with the RScl is governed by the
renormalization group equation (RGE)
\begin{equation}
\frac{ \partial a(\mu^2;c_2,c_3,\ldots)}{\partial \ln \mu^2}
= - \beta_0 a^2 (1 + c_1 a + c_2 a^2 + c_3 a^3 + \cdots) \ ,
\label{RGE}
\end{equation}
where the parameters $c_j\!\equiv\!\beta_j/\beta_0$ ($j \geq 2$) 
characterize the
choice of the RSch, and $\beta_0$ and $c_1\!\equiv\!\beta_1/\beta_0$ 
are universal constants.\footnote{
$\beta_0 = (11 - 2 n_f/3)/4$,
$c_1 = (102 - 38 n_f/3)/(16 \beta_0)$, where $n_f$ is
the number of active quark flavors.}
The evolution of parameter $a$ with $c_j$'s ($j \geq 2$)
is governed by analogous differential equations, which
follow from the Stevenson equation 
(see Appendix A of Ref.~\cite{stevenson}). 
The next--to--leading (NL) coefficient $w_1$ has been
calculated in Ref.~\cite{GLZN}, and the NNL
coefficient $w_2$ in Ref.~\cite{LV}. At the
specific RScl $\mu^2=Q^2$ and ${\overline {\rm MS}}$ RSch,
and when the number of the active quark flavors is $n_f\!=\!3$,
these coefficients have the values\footnote{
The superscript `(0)' in $w_j^{(0)}$
denotes the special RScl--choice $\mu^2\!=\!Q^2$
and ${\overline {\rm MS}}$ RSch.}
\begin{eqnarray}
w_1^{(0)} \equiv  w_1(\mu^2\!=\!Q^2) &=& 3.58333 \ ,
\label{r1o0}
\\
w_2^{(0)}  \equiv  w_2(\mu^2\!=\!Q^2; c_2^{\overline {\rm MS}})
&=& 20.2153 + w_2^{(0)}(l.l.) \ ,
\label{r2o0}
\\
w_2^{(0)}(l.l.)& =&  -1.23954 \ . 
\label{r2o0ll}
\end{eqnarray}
In the NNL coefficient, the small ``light--by--light'' part 
was separated off. The ``light--by--light'' contribution
should not be included in resummations of
$\Delta(Q^2;{\rm P})$, as will be argued at the end of
this Subsection.

\begin{figure}[ht]
 \centering\epsfig{file=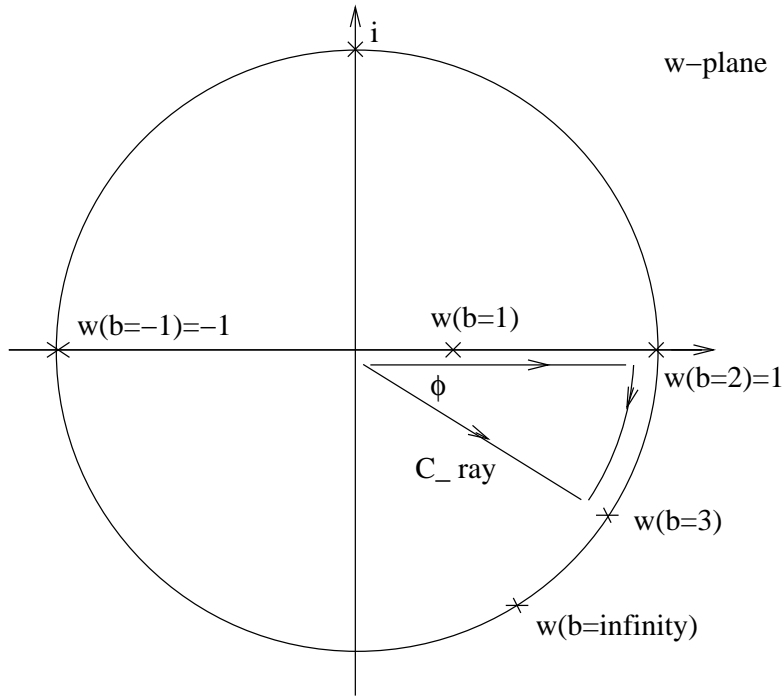}
\vspace{0.3cm}
\caption{\footnotesize
Integration in the $w$--plane
along the ${\cal C}_{-}$ ray $w\!=\!x \exp(-{\rm i} \phi)$
($0\!<\!x\!<\!1$, $\phi\!=\!0.67967$)
gives the same result as the integration parallel to the
positive real axis ($0 < w < 1$) and arc
$w = \exp(-{\rm i} {\phi}^{\prime})$ ($0 < {\phi}^{\prime} < \phi$).
If integrating in the first quadrant, the paths are
those obtained from the presented paths by
reflection across the real axis, and the  ${\cal C}_{+}$ ray
is $w\!=\!x \exp(+{\rm i} \phi)$.}
\label{GLSpath1}
\end{figure}

The Borel integral (\ref{borelintegral2}),
which will be together with (\ref{masterequation})
the basis for our resummation, is independent of
the choice of the RScl $\mu^2$ and the RSch
($c_2, c_3, \ldots$) used in the integrand.\footnote{
The location of the renormalon pole $b=1$ and the
power $\nu=32/81$ (\ref{nu}) are independent of
the choice of the RScl and RSch.}
Thus, we can rewrite it as (see also Fig.~\ref{GLSpath1})
\begin{eqnarray}
\Delta_{\rm P}(Q^2 \pm i \epsilon) & = & 
e^{\pm i \phi} \frac{1}{\beta_0} \int_0^1 dx \frac{db(w)}{dw}
\exp \left[ - \frac{b(w)}{\beta_0 a(\mu^2)} \right]
\frac{R(b(w);\mu^2/Q^2)}{(1 - b(w))^{113/81}}
{\Bigg|}_{w=x e^{ \pm i \phi}} \ ,
\label{BT3}
\end{eqnarray}

At this stage we can, as a byproduct of formula (\ref{rb-new}),
obtain an estimate of the yet unknown ${\rm N}^3{\rm L}$
coefficient $w_3$ appearing in the perturbative
expansion (\ref{cGLSP1}) \cite{lee2,lee7}. If working with the
RScl $\mu^2\!=\!Q^2$ and in the ${\overline {\rm MS}}$
scheme, the expansion of $R(b(w))$ in powers
of $w$ gives us:
\begin{eqnarray}
R(b(w)) & = & 1 + 0.526749 \; w + 0.709369 \; w^2
\nonumber\\
&& + (-43.2574+0.277464 w_3^{(0)}) w^3 + {\cal O}(w^4) \ ,
\label{Rbwexp}
\end{eqnarray}
where we excluded the ``light--by--light'' part (\ref{r2o0ll})
of the NNL coefficient $w_2^{(0)}$ (\ref{r2o0}).
Looking at the coefficients appearing in (\ref{Rbwexp}),
it is reasonable to expect that the coefficient $R_3$
at $w^3$ is $|R_3| \sim 1$. If we assume 
$R_3 = 1 \pm 1$, we obtain a rather stringent
estimate $w_3^{(0)} = 159.5 \pm 3.6$. 
If we adopt a more cautious assumption
$|R_3| \stackrel{<}{\sim} 10^1$, we obtain
$w_3^{(0)} \approx 160 \pm 30$. 
If we apply Pad\'e approximant (PA) $[1/1]_{R}(w)$
to the expansion (\ref{Rbwexp}) and re--expand it
back in powers of $w$ up to $w^3$, we obtain
an estimate $w_3^{(0)} =159.3$. On the other
hand, if applying the $[1/2]_{R}(w)$ and demanding
$w_{\rm pole}\!=\!1$ (i.e., $b\!=\!+2$, ${\rm IR}_2$),
the prediction is $w_3^{(0)}=158.5$; if demanding
$w_{\rm pole}\!=\!-1$ (i.e., $b\!=\!-1$, ${\rm UV}_1$),
the prediction is $w_3^{(0)}=157.0$.
Very similar estimates are obtained if we do not apply
the conformal transformation $b(w)$ (\ref{ct}).
Therefore, we will adopt the following estimate
for $w_3^{(0)}$
\begin{equation}
w_3^{(0)} \ \left[
\equiv w_3(\mu^2\!=\!Q^2; {\overline {\rm MS}})
\right] = 158 \pm 30 \ .
\label{r3est}
\end{equation}
We emphasize that this estimate excludes the
``light--by--light'' contributions, which are
assumed to be suppressed at the ${\rm N}^3{\rm L}$
order. The exclusion of the ``light--by--light''
contributions reduces the (NNL) perturbative
expansion of the (non-polarized) GLS sum rule
to that of the Bjorken polarized sum rule (BjPSR) \cite{LV}.
It is interesting that the 
effective charge method (ECH) \cite{ECH,Kataev:1982gr,Gupta}
and the (TPS) principle of minimal sensitivity (PMS)
\cite{stevenson} predict $w_3^{(0)} \approx 130$
\cite{KS}, based on the assumption that
$c_3^{\rm ECH} \approx c_3^{\rm PMS} \approx 
c_3^{\overline {\rm MS}}$. Further, an
RScl-- and RSch--invariant method \cite{CK1}
that is somewhat related to the PA and PMS approaches,
also predicts $w_3^{(0)} \approx 130$.
This is at the lower end of our new estimate (\ref{r3est}).
We thus conclude that the explicit (and exact) structure of
the leading IR renormalon of the GLS--observable
(BjPSR--observable) in the Borel plane, 
as given in Eq.~(\ref{rb-new}),
is responsible for the somewhat higher estimate
of $w_3^{(0)}$ in comparison with the ECH, PMS,
and PA--related methods of resummation.

We add here a few remarks on the question
of the ``light--by--light'' contributions.
The power expansion of the massless part of the GLS sum rule 
to ${\rm N}^3{\rm L}$
order $\sim\!a^3$ (see Refs.~\cite{GLZN,LV}),
in the $\overline{\rm{MS}}$ scheme and at the RScl
$\mu^2\!=\!Q^2$, is given by
 \begin{eqnarray}
(-\frac{3}{4}C_F) W_{0}(Q^2) &=& a (-\frac{3}{4}C_F)+ a^2 C_F
  (\frac{21}{32}C_F-\frac{23}{16}C_A+\frac{1}{4}n_{f})\nonumber \\ & &
+ a^3 \Bigg[ -\frac{3}{128}C^3_F +
(\frac{1241}{576}-\frac{11}{12}\zeta_3)C^2_FC_A +
(-\frac{5437}{864}+\frac{55}{24}\zeta_5)C_FC^2_A \nonumber \\ & &
+ (-\frac{133}{1152}-\frac{5}{24}\zeta_3)C^2_Fn_{f} +
(\frac{3535}{1728}+\frac{3}{8}\zeta_3-\frac{5}{12}\zeta_5)C_{F}
C_{A}n_{f} -\frac{115}{864}C_{F}n^{2}_{f}\nonumber \\ & &
+n_{f}\frac{d^{abc}d^{abc}}{N_{c}}(-\frac{11}{192}
+\frac{1}{8}\zeta_3)\Bigg] + {\cal O}(a^4) \ ,
\label{lbyl}
\end{eqnarray}
where the Casimir coefficients for QCD ($N_c\!=\!3$) are
$C_F\!=\!4/3$, $C_A\!=\!3$, and the group--theoretical
factor appearing in the last term in (\ref{lbyl}) is
\begin{equation}
d^{abc}d^{abc}= \frac{40}{3} \ .
\label{Casimir}
\end{equation}
This group--theoretical factor is not present in the
calculation of the GLS sum rule up to the two--loop ($\sim\!a^2$) order,
and it appears for the first time at the three--loop ($\sim\!a^3$) order. 
This term is called ``light--by--light,'' it corresponds 
to diagrams with a new topology involving exchange of three gluons. 
In our calculation we will add this ``light--by--light'' 
contribution $\Delta_{l.l.}(Q^2)$, given by the last term in
Eq.~(\ref{lbyl}), as a separate term not included in our
resummation approach. The reason for this is the following:
Resummation approaches cannot be expected to predict (and resum)
those higher order terms which are characterized by 
new higher order group--theoretical factors, when we have
only one such term ($\sim\!a^3$) explicitly available.
We assume that such resummed ``light--by--light'' contributions
are small, comparable to the quite small $\sim\!a^3$ 
``light--by--light'' contribution in (\ref{lbyl}).
Similar considerations can be found in 
Refs.~\cite{LbL1,KS}, in cases of various observables
and beta functions.

\subsection{Inclusion of the massive quark flavor
(\lowercase{$c$}) contribution and nuclear corrections}
\label{massive}

In the approximation of massless quarks, 
the calculation of the ${\rm N}^3{\rm L}$ ($\sim\!a^3$)
QCD correction to the GLS sum rule 
has been carried out in Ref.~\cite{LV}.
The  inclusion of the heavy ($c$--)quark contributions is 
important at the precision level at which we are working.
In addition, it is important in order to estimate 
the scales $Q^2$ where the ($c$--)quarks can be treated as massless
or massive quarks. This point is important because it indicates 
which number of light flavors $n_f$ should be used in the (resummed)
``massless'' part of the the perturbation series. 
The calculation of the heavy flavor contribution to the 
GLS sum rule up to the second order ($\sim\!a^2$)  
was performed in Ref.~\cite{Buza:1996xr} and 
discussed in \cite{Blumlein:1998sh}. 
According to this approach: (a) the quarks $u,d,s$ are massless 
and result in the dominant $n_f\!=\!3$ massless QCD contribution 
to the GLS sum rule $\Delta(Q^2)$;
(b) for $Q^2 \approx 2$--$4 \ {\rm GeV}^2$, the massive
flavor is the $c$--quark and it contributes a relatively
small correction to the aforementioned $n_f\!=\!3$
massless contribution.
The other heavy quark flavors are ignored due to the
strong suppression by the mixing angles in the 
Cabibbo--Kobayashi--Maskawa matrix and by the small
values of $Q^2$.

The heavy ($c$--)flavor
correction contributions to $\Delta(Q^2)$ are
\begin{eqnarray}
\Delta_c(Q^2) &=& \Delta_c^{(1)} + \Delta_c^{(2)} \ ,
\label{cGLScdef}
\\
\Delta _{c}(Q^2)^{(1)}&=&\left[ 
\frac{1}{3(1+\xi) } - a(\mu^2) \frac{C_F}{4}
\left\{ \frac{1}{1+\xi }+
2 \frac{\ln (1+\xi )}{1+\xi }\right\} \right] \sin ^{2}\theta _{c} \ ,
\label{mc1}
\\
\Delta _{c}(Q^2)^{(2)}&=&  - a(\mu^2)^2 \frac{C_{F}\,T_{F}}{16}
\Bigg[ \left(
\frac{1}{105}\xi ^{2}+\frac{16}{45}\xi \right) \ln \xi \nonumber
\\ 
& & +\frac{1}{\lambda ^{4}}\left( \frac{2}{105}\xi
+\frac{2783}{ 315}+\frac{6740}{63}\frac{1}{\xi
}+\frac{137552}{315}\frac{1}{\xi ^{2}}+
\frac{62528}{105}\frac{1}{\xi ^{3}}\right)    \, \nonumber \\ 
& & -\frac{1}{\lambda ^{5}} \Bigg( \frac{1}{105}\xi
^{2}+\frac{142}{315}\xi +\frac{494}{63}+\frac{1516}{21}
\frac{1}{\xi }+\frac{23024}{63}\frac{1}{\xi
^{2}}+\frac{298432}{315}\frac{1}{ \xi ^{3}}\nonumber \\ 
& & + \frac{102656}{105}\frac{1}{\xi ^{4}} \Bigg) \ln (\frac{\lambda
+1}{ \lambda -1})-\frac{20}{3}\frac{1}{\xi ^{2}}
\ln^{2}(\frac{\lambda +1}{ \lambda -1}) \Bigg] \ ,
\label{mc2}
\end{eqnarray}
where $\xi\!=\!Q^2/m^2_c$, $\lambda = \sqrt{1\!+\!4 m_c^2/Q^2}$, 
$C_F\!=\!4/3$, $T_F\!=\!1/2$.
The heavy flavor corrections (\ref{mc1})--(\ref{mc2})
will not be included in the resummation procedure
for $\Delta$, but will be added separately.

In addition to the discussed heavy flavor
contributions, there are also nuclear corrections
to the GLS sum rule, due to the 
nuclear effects in the $F_3$ structure functions. 
These effects were calculated in Ref.~\cite{kulagin} and 
were found for the iron target (used by the CCFR
Collaboration) to be small:
\begin{equation} 
\Delta_{\rm Fe}(Q^2) \approx \frac{4 \cdot 10^{-3} \ {\rm GeV}^2}{Q^2}
\ .
\label{cGLSFe}
\end{equation}
This contribution, at $Q^2=2$--$3 \ {\rm GeV}^2$, 
turns out to be in its magnitude smaller
than the heavy flavor effects by a factor of 3--4,
but larger than the ``light--by--light'' contribution
by a factor of about 2.

\subsection{Extraction of \lowercase{$\alpha_s$} values}
\label{extract}

The values of the GLS sum rule $M_3(Q^2)$, for
various specific $Q^2$, have been obtained from
the experimentally extracted values of the structure 
function $F_{3}(x;Q^2)$, by the Fermilab CCFR
Collaboration \cite{ccfr}. They have large
systematic experimental uncertainties, primarily
because of the uncertainties in the normalization
of $x F_3$ and in the integration in the regions
$x \ll 1$ and $x > 0.5$. The systematic uncertainties
are somewhat smaller, comparable to the statistical
experimental uncertainties, only when the
values of the exchanged boson virtuality are 
$Q^2\!=\!2.00, 3.16 \ {\rm GeV}^2$ (see their Table III). 
These values, including the target mass correction terms
of (\ref{M3o1})--(\ref{M3o2}), as well as the nuclear
correction term (\ref{cGLSFe}), are\footnote{
In the entries of Table III of Ref.~\cite{ccfr}, the target mass
corrections are included, but the nuclear corrections
neglected.}
\begin{eqnarray}
Q^2 = 2.00 \ {\rm GeV}^2 \ : \quad
M_3(Q^2) &=& 2.49 \pm 0.13 \quad \Leftrightarrow \
\Delta(Q^2) = 0.168 \pm 0.043 \ ,
\label{exp2}
\\
Q^2 = 3.16 \ {\rm GeV}^2 \ : \quad
M_3(Q^2) &=& 2.55 \pm 0.12 \quad \Leftrightarrow \
\Delta(Q^2) = 0.149 \pm 0.039 \ .
\label{exp3}
\end{eqnarray}
Here we added in quadrature the statistical and the
systematic experimental uncertainties.
In addition, the values $\Delta(Q^2)$ in
(\ref{exp2})--(\ref{exp3}) have the aforementioned
nuclear correction contribution (\ref{cGLSFe}) subtracted out;
however, $\Delta(Q^2)$ still includes the massless
(perturbative and nonperturbative) contribution (\ref{masterequation}),
the light--by--light contribution $w_2(l.l.)\;a^3$ [cf.~(\ref{r2o0ll})]
and the $c$--quark contribution (\ref{cGLScdef})--(\ref{mc2}).

For a given value of $\alpha_s(Q^2,{\overline {\rm MS}})$,
our expression for $\Delta(Q^2) =
\Delta_{\rm P}(Q^2)\!+\!\Delta_{\rm NP}(Q^2)\!+\!\Delta_{l.l.}(Q^2)\!+\!
\Delta_{c}(Q^2)$,
given in the previous Subsections, still has the freedom (ambiguity)
of the choice of the RScl $\mu^2$ and the RSch. 

In the following analysis, we fix first the RSch
to be ${\overline {\rm MS}}$, with the beta function
being the PA $\beta(x) = [2/3](x)$ based on the
${\rm N}^3{\rm L}$ TPS of the $\beta_{\overline {\rm MS}}(x)$.
This PA choice of the beta function has reasonable
behavior in the region of large $x\equiv \alpha_s/\pi$
and has been used in Refs.~\cite{lee2,lee3,CK1}
in the analyses of low energy QCD observables.
We will comment later on how the results change
when we use ${\rm N}^3{\rm L}$ TPS of the 
$\beta_{\overline {\rm MS}}(x)$, and when we
change the RSch even more drastically.

\begin{figure}[ht]
 \centering\epsfig{file=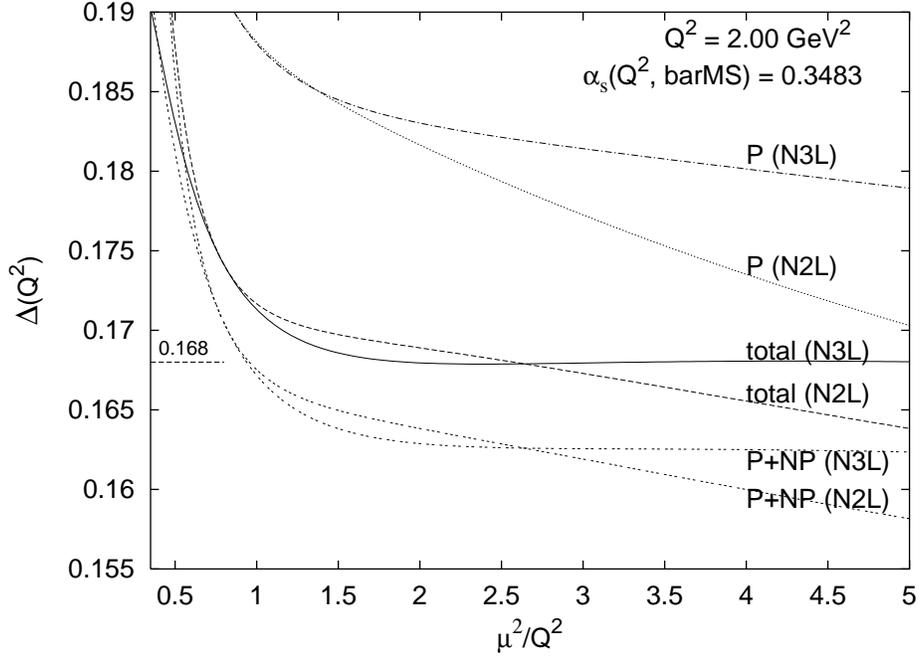}
\vspace{0.3cm}
\caption{\footnotesize The (total) 
$\Delta(Q^2) =
\Delta(Q^2,{\rm P}\!+\!{\rm NP}) + \Delta(Q^2,l.l.) +
\Delta(Q^2, c-{\rm quark})$, as well as the
$n_f\!=\!3$ perturbative part $\Delta(Q^2,P)$
and nonperturbative part $\Delta(Q^2,NP)$,
as functions of the renormalization
scale $\mu^2$, as given by the applied
resummation method. Given are the results
at the ${\rm N}^3{\rm L}$ level,
and for comparison, at the NNL level.
The curves are for $Q^2 = 2.00 \ {\rm GeV}^2$
and $\alpha_s(Q^2,{\overline {\rm MS}}) = 0.3483$.
}
\label{xi2}
\end{figure}

\begin{figure}[ht]
 \centering\epsfig{file=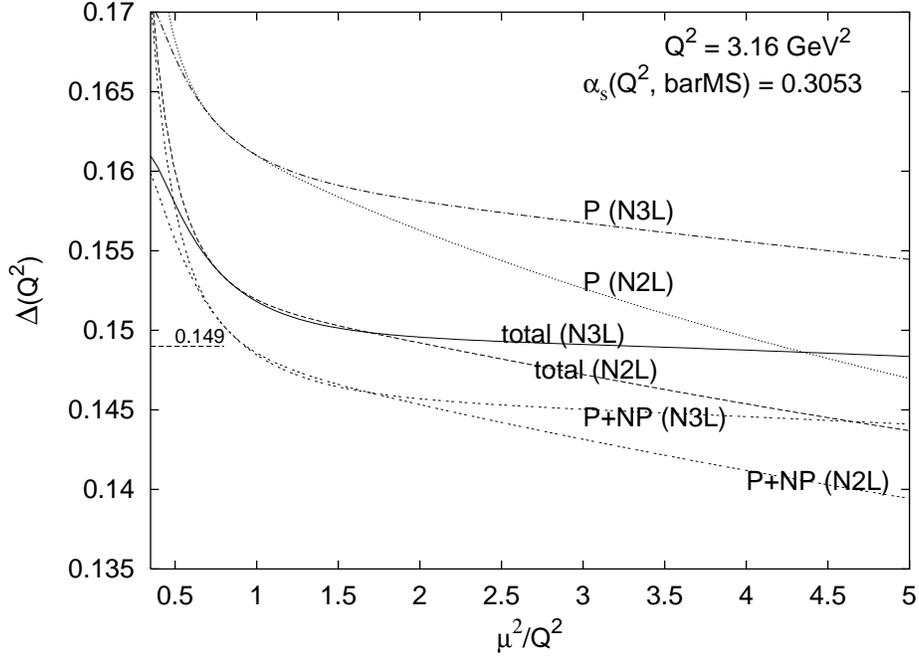}
\vspace{0.3cm}
\caption{\footnotesize Same as in Fig.~\ref{xi2},
this time for $Q^2 = 3.16 \ {\rm GeV}^2$
and $\alpha_s(Q^2,{\overline {\rm MS}}) = 0.3053$.
}
\label{xi3}
\end{figure}

The RScl--dependence is yet another source of
the theoretical ambiguity. In Figs.~\ref{xi2}
and \ref{xi3}
we show the dependence of the predicted values
of $\Delta(Q^2)$ on the RScl $\mu^2$
at the NNL and ${\rm N}^3{\rm L}$ level,
at $Q^2\!=\!2.00, 3.16 \ {\rm GeV}^2$, respectively,
for given fixed representative values of
$\alpha_s(Q^2,{\overline {\rm MS}})$.
The NNL level means that we take for
$R(b(w))$ in the Borel integrations
in (\ref{BT3}) and (\ref{masterequation}) the NNL TPS
(quadratic polynomial in $w$),
i.e., only the knowledge of $w_1$ and $w_2$
coefficients is used. The ${\rm N}^3{\rm L}$ level
means that $R(b(w))$ is the ${\rm N}^3{\rm L}$ TPS 
(cubic polynomial in $w$), i.e., we use, in addition,
$w_3^{(0)} = 158$ according to the estimate (\ref{r3est}).
From Figs.~\ref{xi2},\ref{xi3} we see that the
${\rm N}^3{\rm L}$ expressions drastically
improve the stability of the predictions under
the change of the RScl. In the case 
$Q^2\!=\!2.00 \ {\rm GeV}^2$, the ${\rm N}^3{\rm L}$ values
of the massless quantity $\Delta_{{\rm P}\!+\!{\rm NP}}(Q^2)$ 
achieve minimal $\ln \mu^2$--sensitivity at 
$\mu^2/Q^2 \approx 3.3$~: 
$d \Delta_{{\rm P}\!+\!{\rm NP}}(Q^2)/d \ln \mu^2 \approx - 1.45 \cdot
10^{-4}$.
At $Q^2 = 3.16 \ {\rm GeV}^2$, there is unfortunately
no point of minimal RScl--sensitivity; the slope is
negative and getting weaker when $\mu^2$ increases. However,
the RScl--sensitivity is very weak when 
$\mu^2/Q^2 > 2.5$, and almost stabilizes when
$\mu^2/Q^2 \approx 3.3$~: 
$d \Delta_{{\rm P}\!+\!{\rm NP}}(Q^2)/d \ln \mu^2 \approx - 1.55 \cdot
10^{-3}$.
Therefore, we will take, for definiteness, the RScl choice 
$\mu^2 = 3.3\;Q^2$
in the case of both $Q^2$ values. Later on, we will comment
on the ambiguity of our results when the RScl is varied.

\begin{figure}[ht]
 \centering\epsfig{file=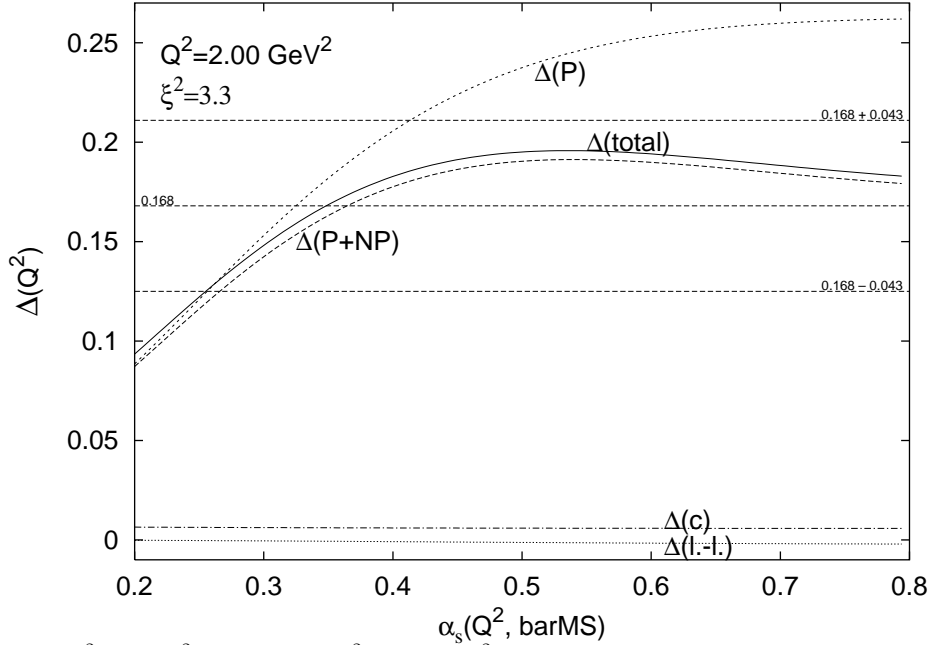}
\vspace{0.cm}
\caption{\footnotesize The (total) 
$\Delta(Q^2) =
\Delta(Q^2,{\rm P}\!+\!{\rm NP}
) + \Delta(Q^2,l.l.) +
\Delta(Q^2, c-{\rm quark})$,
and the separate parts, as functions of
$\alpha_s(Q^2,{\overline {\rm MS}})$,
as given by the applied resummation method.
The renormalization scale was fixed to
be $\xi^2 \equiv \mu^2/Q^2 = 3.3$,
and the $W$--boson virtuality is $Q^2 = 2.00 \ {\rm GeV}^2$.
The present experimental bounds and the
central value, for the total $\Delta(Q^2)$
in this case, are denoted as three horizontal dashed lines.
}
\label{al2}
\end{figure}

\begin{figure}[ht]
 \centering\epsfig{file=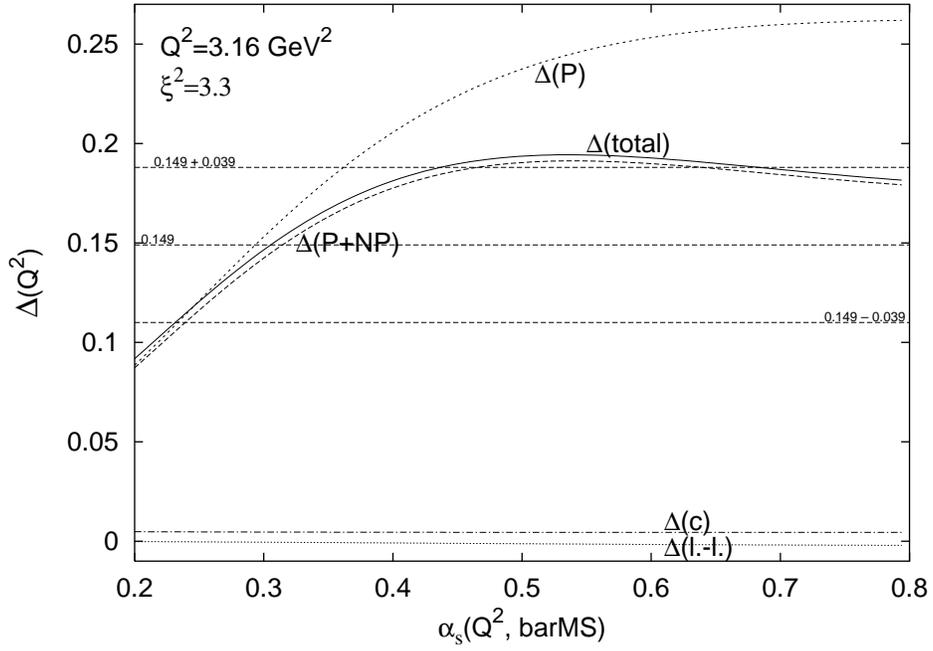}
\vspace{0.3cm}
\caption{\footnotesize Same as in Fig.~\ref{al2},
but this time for the $W$--boson virtuality
$Q^2 = 3.16 \ {\rm GeV}^2$.
}
\label{al3}
\end{figure}

Having fixed the RSch (${\overline {\rm MS}}$, with
$[2/3]_{\beta}(x)$ PA) and the RScl ($\mu^2 = 3.3\;Q^2$),
our expressions for $\Delta(Q^2) =
\Delta(Q^2,{\rm P}\!+\!{\rm NP}\!+\!l.l.\!+\!c)$
become unambiguous functions of the input value of 
$\alpha_s(Q^2;{\overline {\rm MS}})$.
Adjusting simply the latter value so that the
experimental values (\ref{exp2})--(\ref{exp3})
are achieved, we obtain:
\begin{eqnarray}
Q^2 = 2.00 \ {\rm GeV}^2 \ : \quad
\alpha_s(Q^2;{\overline {\rm MS}}) & = &
0.348^{+ ?}_{-0.093} (\rm exp.)
\label{al2o1}
\\
Q^2 = 3.16 \ {\rm GeV}^2 \ : \quad
\alpha_s(Q^2;{\overline {\rm MS}}) & = &
0.305^{+0.132}_{-0.074} (\rm exp.)
\label{al3o1}
\end{eqnarray}
The central, upper and lower values here correspond
to the pertaining experimental values of $\Delta(Q^2)$
in (\ref{exp2})--(\ref{exp3}). 
In the case $Q^2 = 2.00 \ {\rm GeV}^2$, the upper bound 
for the coupling parameter cannot be obtained from the experimental 
data because the applied resummation method predicts 
values of $\Delta(Q^2)$ which are always lower
than the presently allowed experimental upper bound:
$\Delta(Q^2)_{\rm meth.} \leq 0.196$ 
[$ < \Delta(Q^2)_{\rm max \ exp.} = 0.168\!+\!0.043$].
The situation in the case $Q^2 = 3.16 \ {\rm GeV}^2$ is
somewhat similar, the experimental upper bound being now
slightly below the maximal value allowed by the method.
This situation is presented graphically in
Figs.~\ref{al2} and \ref{al3} which show $\Delta(Q^2)$
as function of $\alpha_s(Q^2)$
as predicted by the applied method, for 
$Q^2 = 2.00, 3.16 \ {\rm GeV}^2$ cases, respectively.
The central, upper and lower experimental bounds are
included as straight dotted lines.

We now return to the question of the uncertainties of
our predictions under the variation of the RScl
and RSch. If we vary the RScl parameter
$\xi^2 \equiv \mu^2/Q^2$ around 3.3
across the interval $1.5 \leq \xi^2 \leq 5$,
the predictions for $\alpha_s(Q^2;{\overline {\rm MS}})$
vary by at most $\pm 0.002$.
On the other hand, variation of the RSch leads
to larger ambiguities.
For example, if we repeat the calculation in
the `t Hooft RSch ($c_2 = c_3 = 0$,
we take $[2/3]_{\beta}(x)$ for definiteness),\footnote{
For comparison, $c_2^{\overline {\rm MS}} = 4.471$
$c_3^{\overline {\rm MS}} = 20.99$, for
$n_f\!=\!3$.} 
keeping $\xi^2\!=\!3.3$,
the predictions for $\alpha_s(Q^2;{\overline {\rm MS}})$
change by $0.009$ ($Q^2\!=\!2.00 \ {\rm GeV}^2$)
and $0.005$ ($Q^2\!=\!3.16 \ {\rm GeV}^2$).
We will regard these values as
characteristic values for the RSch--uncertainties
of our results. The replacement of the
${\overline {\rm MS}}$ $[2/3]_{\beta}(x)$ 
by the ${\rm N}^3{\rm L}$ TPS ${\overline {\rm MS}}$
beta function changes our predictions for 
$\alpha_s(Q^2;{\overline {\rm MS}})$ by only
about $0.001$.

Yet another source of the theoretical uncertainty
in our predictions may be the truncation in the
(${\rm N}^3{\rm L}$) TPS $R(b(w))$.
We regard the uncertainty $\pm 30$ in the
estimated value of $w_3^{(0)}$, Eq.~(\ref{r3est}),
as the major source of the truncation uncertainty.
This changes the prediction for $\alpha_s(Q^2;{\overline {\rm MS}})$
by only $\pm 0.001$. 

The other sources of theoretical uncertainty
come from the massive ($c$--)quark contributions
presented in Subsec.~\ref{massive}, due
to the uncertainties $m_c = 1.25 \pm 0.10 \ {\rm GeV}$
and $\sin(\theta_{\rm Cabibbo}) = |V_{cd}| = 0.224 \pm 0.016$
\cite{PDG2000}. The resulting uncertainties of the
predictions for $\alpha_s(Q^2;{\overline {\rm MS}})$
are $\pm 0.002$ and $\pm 0.002$, respectively,
when $Q^2\!=\!2.00 \ {\rm GeV}^2$,
and $\pm 0.001$ and $\pm 0.001$ when
$Q^2\!=\!3.16 \ {\rm GeV}^2$.

The final predictions for $\alpha_s$
are presented in Table \ref{tabl1}. 
\begin{table}[ht]
\par
\begin{center}
\begin{tabular}{l| c c| c c}
   & $Q^2=2.00 \ {\rm GeV}^2$ & $\Rightarrow \ M^2_Z$ &
$Q^2=3.16 \ {\rm GeV}^2$ & $\Rightarrow \ M^2_Z$ \\  
\hline \hline
$\alpha_s(Q^2)$ & $0.348$ & $0.1166$ & $0.305$ & $0.1167$ \\
\hline
$\delta \alpha_s$ ($\!>\!0$, exp.) & 
$+ ?$    & $+ ?$    & $+0.132$ & $+0.0128$ \\
$\delta \alpha_s$ ($\!<\!0$, exp.) & 
$-0.093$ & $-0.0115$& $-0.074$ & $-0.0118$ \\
$\delta \alpha_s$ (RSch) & 
$\pm 0.009$ & $\pm 0.0008$ & $\pm 0.005$ & $\pm 0.0006$ \\
$\delta \alpha_s$ (RScl) & 
$\pm 0.002$ & $\pm 0.0002$ & $\pm 0.002$ & $\pm 0.0003$ \\
$\delta \alpha_s$ ($w_3$) &
$\pm 0.001$ & $\pm 0.0001$ & $\pm 0.001$ & $\pm 0.0002$ \\
$\delta \alpha_s$ ($m_c$) &
$\pm 0.002$ & $\pm 0.0002$ & $\pm 0.001$ & $\pm 0.0002$ \\
$\delta \alpha_s$ ($\sin \theta_c$) &
$\pm 0.002$ & $\pm 0.0003$ & $\pm 0.001$ & $\pm 0.0002$ \\
 $\delta \alpha_s$ (evol.) &
 -- &  $\pm 0.0003$ & --- & $\pm 0.0003$ 
\end{tabular}
\end{center}
\caption {\footnotesize Predictions for 
${\alpha}_s^{{\overline {\rm MS}}}(Q^2)$ and
${\alpha}_s^{{\overline {\rm MS}}}(M^2_Z)$,
extracted by the comparison of the results of
the applied resummation method with
the measured GLS values (\ref{exp2}),(\ref{exp3}).}
\label{tabl1}
\end{table}
In the Table, we presented the central predictions for
$\alpha_s(Q^2;{\overline {\rm MS}})$
when $Q^2\!=\!2.00$ and $3.16 \ {\rm GeV}^2$,
and the uncertainties of the predictions due to
various sources. Further, we RGE--evolved these
predictions to the canonical scale $M_Z^2$
and included the results in Table \ref{tabl1}.
This RGE evolution was carried out by using the
$[2/3]$ Pad\'e approximant of the four--loop
${\overline {\rm MS}}$ TPS beta function, using the values of
the four--loop coefficient $c_3(n_f)$ \cite{RVL}
and the corresponding three--loop matching
conditions \cite{Chetyrkinetal} for the flavor thresholds.\footnote{
For details on the corresponding evolution
uncertainties, we refer to Ref.~\cite{lee2}. They include the
variation when the [2/3] Pad\'e form of the beta function is replaced
by the TPS form.} 
When adding in quadrature the various
theoretical uncertainties, the predictions of Table \ref{tabl1}
can be summarized as
\begin{eqnarray}
Q^2\!=\!2.00 \ {\rm GeV}^2: \quad
\alpha_s(Q^2;{\overline {\rm MS}}) & = &
0.348^{+ ?}_{-0.093} (\rm exp.)  \pm 0.010 (\rm th.) \ , 
\label{al2o2}
\\
\Rightarrow \quad 
\alpha_s(M^2_Z;{\overline {\rm MS}}) &=& 
0.1166^{+ ?}_{-0.0115} (\rm exp.) \pm 0.0010 (\rm th.) \ ;
\label{al2o3}
\end{eqnarray}
and
\begin{eqnarray}
Q^2\!=\!3.16 \ {\rm GeV}^2: \quad
\alpha_s(Q^2;{\overline {\rm MS}}) & = &
0.305^{+0.132}_{-0.074} (\rm exp.) \pm 0.006 (\rm th.) \ ,
\label{al3o2}
\\
\Rightarrow \quad
\alpha_s(M^2_Z;{\overline {\rm MS}}) &=& 
0.1167^{+0.0128}_{-0.0118} (\rm exp.) \pm 0.0008 (\rm th.) \ .
\label{al3o3}
\end{eqnarray}
We see that the predictions of the applied method suggest
that the experimental data on the GLS
should be refined significantly in order 
to increase the predictive power for the QCD coupling parameter. 

Another observation, evident from Figs.~\ref{xi2}--\ref{al3},
is that the nonperturbative massless contributions (NP)
to $\Delta(Q^2)$ are very significant, and negative.
They have their origin, as explained in the previous Section,
in the first IR renormalon singularity at $b\!=\!1$,
or equivalently, they correspond approximately 
to the $d\!=\!2$ massless power corrections ($\sim 1/Q^2$).
For $Q^2\!=\!2.00, 3.16 \ {\rm GeV}^2$,
they lead to about $11 \%$, $8 \%$, decrease of the
value of $\Delta(Q^2)$, respectively. 
This is to be contrasted
with the heavy ($c$--)quark contribution\footnote{
The latter is, to a large degree, a $d\!=\!2$ massive power
correction $\propto\!m_c^2/Q^2$, see Subsec.~\ref{massive}.}
which is positive and leads to only about $3.6 \%$,
$3.1 \%$ increase of $\Delta(Q^2)$, respectively.
If the $c$--quark contribution were not included,
the central predicted values in (\ref{al2o2})--(\ref{al3o3})
would change to $\alpha_s(Q^2)=0.367$, $0.316$ 
[$\alpha_s(M^2_Z)=0.1183$, $0.1180$] for
$Q^2\!=\!2.00$, $3.16 \ {\rm GeV}^2$, respectively.

The small negative ``light--by--light'' contribution was
separated from our resummation and then added as the term 
$\Delta_{l.l.}(Q^2) \approx 
w_2^{(0)}(l.l.) a^3(\mu^2;c_2,\ldots)$.
The ``light--by--light'' part
of $\Delta_{l.l.}(Q^2)$, by the special topology 
of the Feynman diagrams representing it, is a quasiobservable
in the sense that it is RScl-- and RSch--invariant.
Thus, the coefficient $w_2^{(0)}(l.l.)$ (\ref{r2o0ll})
is the leading coefficient of that quasiobservable
and is therefore unchanged under the changes of the
RScl and RSch. The ``light--by--light'' part decreases
$\Delta(Q^2)$ by only about $0.4 \%$ and
$0.3 \%$, for $Q^2 = 2.00, 3.16 \ {\rm GeV}^2$,
respectively.

\section{Comparison with other approaches}
\label{comparison}

One may ask how crucial is the
introduction of the conformal transformation
(\ref{ct}) for obtaining the numerical predictions
(\ref{al2o2})--(\ref{al3o3}). If we repeat the
same analysis, but this time without the conformal
transformation, and keeping $\xi^2\!=\!3.3$, we obtain
\begin{eqnarray}
Q^2\!=\!2.00 \ {\rm GeV}^2 : \
\alpha_s(Q^2;{\overline {\rm MS}}) & = &
0.346^{+ ?}_{-0.092} (\rm exp.), \
\alpha_s(M_Z^2;{\overline {\rm MS}}) = 
0.1163^{+?}_{-0.0113} (\rm exp.);
\label{al2nct}
\\
Q^2\!=\!3.16 \ {\rm GeV}^2 : \
\alpha_s(Q^2;{\overline {\rm MS}}) & = &
0.304^{+0.125}_{-0.073} (\rm exp.)), \
\alpha_s(M_Z^2;{\overline {\rm MS}}) = 
0.1165^{+0.0124}_{-0.0117} (\rm exp.).
\label{al3nct}
\end{eqnarray}
These results are very close to the results (\ref{al2o2})--(\ref{al3o3}).
Thus, we see that the introduction
of the conformal transformation (\ref{ct}), which had the task
of reducing the influence of the UV and the nonleading IR renormalons, 
does not influence significantly the predictions. Therefore, we
can conclude that these renormalon singularities
(at $b\!=\!-1, -2, \ldots$ and $b\!=\!2,3,\ldots$) 
are in GLS numerically much less important than the leading
IR renormalon singularity (at $b\!=\!1$), even when
no conformal transformation is introduced.

We can ask how our predictions compare with
those of other, alternative, OPE based methods which, 
in contrast with the method applied here,
do not take into account explicitly the
structure of the leading IR renormalon singularity 
in the Borel plane. 

One such an alternative method is the (TPS) PMS optimization
of the perturbative contribution, which fixes the
RScl and RSch in the TPS in a judicious manner \cite{stevenson}.
Resummations of the GLS sum rule based on this method  were theoretically
and numerically investigated in 1992 by the authors
of Ref.~\cite{Chyla:1992cg}. They were confronting
the TPS results with the measured values,
paying particular attention to the
RScl-- and RSch--dependence of the
predicted values of $\alpha_s(M_Z^2;{\overline {\rm MS}})$.
For the nonperturbative massless (twist--four) $d\!=\!2$
contribution, they employed the positive value as obtained
in Ref.~\cite{braun-ko} 
[$\Delta_{\rm NP}(Q^2) \approx 0.1 \ {\rm GeV}^2/Q^2$].
Further, the authors of Ref.~\cite{Chyla:1992cg}
accounted for the quark mass threshold
effects ($n_f^{\rm eff.}$) by introducing a judiciously
chosen weighted average of $n_f=\!3\!,4,5$.\footnote{
This is different from our approach, where we 
separately added the contributions 
(\ref{cGLScdef})--(\ref{mc2}) of the heavy ($c$--)quark
as corrections to the massless $n_f\!=\!3$ GLS sum rule,
as recently suggested in Ref.~\cite{Blumlein:1998sh}.}
Furthermore, they used
the GLS measured values available at that time
[$\Delta(Q^2\!=\!3{\rm GeV}^2) = 0.167 \pm 0.027$]
which differ significantly from the presently 
available values (\ref{exp3}). Their central value
prediction was $\alpha_s(M_Z^2)\!=\!0.115$, which is
lower than our central value
predictions (\ref{al2o3}),(\ref{al3o3}).

The CCFR group \cite{ccfr,jhkim} carried out a numerical
analysis similar to that of the authors of Ref.~\cite{Chyla:1992cg},
but with the newer, lower, experimental data
(\ref{exp2})--(\ref{exp3}) for $\Delta(Q^2)$,
and using in the TPS--part ${\overline {\rm MS}}$ RSch
(and RScl $\mu^2\!=\!Q^2$).
Their central value is $\alpha_s(3 \ {\rm GeV}^2)\!=\!0.278$
\cite{jhkim}\footnote{
The values of $\alpha_s(3 \ {\rm GeV}^2)$ obtained from
our central values of Eqs.~(\ref{al2o2}) and (\ref{al3o2}) are
$0.309$ and $0.310$, respectively.}
and $\alpha_s(M_Z^2) = 0.114$,
thus slightly lower than that of Ref.~\cite{Chyla:1992cg},
and significantly lower than our central value 
predictions (\ref{al2o3}) and (\ref{al3o3}). 
The principal reason for this difference shall be discussed
in the following section. Further, if they included in their
method the ${\rm N}^3 {\rm LO}$ term in the TPS, with
$w_3^{(0)}$ as estimated in Eq.~(\ref{r3est}), the predicted
value of $\alpha_s(M^2_Z)$ would decrease by about $0.002$.

Furthermore, the CCFR group mentioned
that their central value increases to 
$\alpha_s(3 \ {\rm GeV}^2)\!\approx\!0.305$ \cite{jhkim}
and $\alpha_s(M^2_Z)\!=\!0.118$ \cite{ccfr,jhkim}\footnote{
We note that the RGE evolution of $\alpha_s$ from $Q^2$ to
$M_Z^2$ gives in our approach different results:
$\alpha_s(3 \ {\rm GeV}^2)\!=\!0.278 (0.305) $ gives
$\alpha_s(M^2_Z)\!=\!0.1124 (0.1162)$ when using the
three--loop or four--loop TPS $\beta$--function, 
$0.1123 (0.1161)$ when using the (four--loop) 
$[2/3]$ PA $\beta$--function -- we use the corresponding
two--loop and three--loop matching conditions for the flavor
thresholds \cite{Chetyrkinetal}; other details
given in Ref.~\cite{lee2}. The CCFR Collaboration
apparently uses an approximate three--loop
RGE evolution [a truncated expansion in inverse powers
of $\ln (\mu^2/{\Lambda}_{\overline {\rm MS}})$]
and different matching conditions for the flavor thresholds, 
possibly of Ref.~\cite{Marciano:1983pj},
giving them the values $0.114 (0.118)$. } 
when they set the twist--four ($d\!=\!2$) contribution
approximately equal to zero. Such higher--twist values for the GLS 
sum rule are suggested by the calculation by the authors
of Ref.~\cite{Dasgupta:1996hh} based on an IR renormalon model with
dispersive approach of Ref.~\cite{Dokshitzer:1995qm}, and also
by the calculation by the authors of Ref.~\cite{Fajfer:fw} 
based on the bag model. It looks reasonable that
the result with the aforementioned IR renormalon model 
method gives prediction rather close to our prediction,
since our method also accounts for the IR renormalon
contribution, although in a different manner. However,
the calculations in Refs.~\cite{Dasgupta:1996hh,Fajfer:fw}
apparently do not give us a clear handle on how to treat
the perturbative part, i.e., whether to take it as a
LO, NLO, or NNLO TPS, or in any other form. This is in contrast
with our method, where the perturbative and nonperturbative
parts are clearly connected with each other. We
discuss this aspect in more detail in the next section.

At this point, we would like to point out that the PMS method has a 
signal casting doubts on its applicability in the discussed GLS cases--
namely, the PMS RScl is very low in this case: 
$\mu^2_{\rm PMS}\approx0.203Q^2$.  For $Q^2=2.00, 3.16 \,{\rm GeV}^2$,
this implies the scales $\mu_{\rm PMS}\approx 0.64,0.80\,{\rm GeV}$, 
respectively, which may be too low for the application of perturbative
approaches such as PMS. The same problem appears when applying
the effective charge (ECH) method \cite{ECH,Kataev:1982gr,Gupta}
to these GLS cases.

The present world average for the QCD coupling parameter is 
${\alpha}_s^{{\overline {\rm MS}}}(M_{\text{z}}^2) =
0.1173 \pm 0.0020$ by Ref.~\cite{Hinchliffe:2000yq} and
$0.1184 \pm 0.0031$ by Ref.~\cite{Bethke:2000ai}.
Comparing this with our predictions
(\ref{al2o3}) and (\ref{al3o3}),
we see that the method applied in the present paper
gives us the central values which agree well
with the present world average. We wish to point
out that this agreement suggests 
that the method applied in the present paper
for the nonperturbative massless correction
to $\Delta(Q^2)$ is at least consistent with the experiments.
If this correction were zero,
or had the opposite sign, the
obtained central prediction for $\alpha_s(M_Z^2)$
would be at the lower edge or even outside
the interval of the present world average.
These considerations do not necessarily
imply, but  indicate, that the applied method gives 
the correct nonperturbative contributions. For
a more definite statement in this respect, the
experimental uncertainties in the GLS sum rule
would have to be reduced significantly.

\section{Borel resummation versus OPE approach}
\label{versus}

In the discussions so far, we considered the amplitudes
of Borel resummation or OPE approach only at fixed values
of $Q^2$. Here we mean, for convenience, by the OPE approach the usual 
perturbative expansion plus a power suppressed term
representing the twist--four contribution.
In this  section we consider them over a continuous 
range of $Q^2$.
This slight change of view
will reveal the characteristic features of the two approaches,
and  enable us to better understand the cause of the
significant difference in the extracted strong coupling constants
seen in the previous section.

\begin{figure}[ht]
 \centering\epsfig{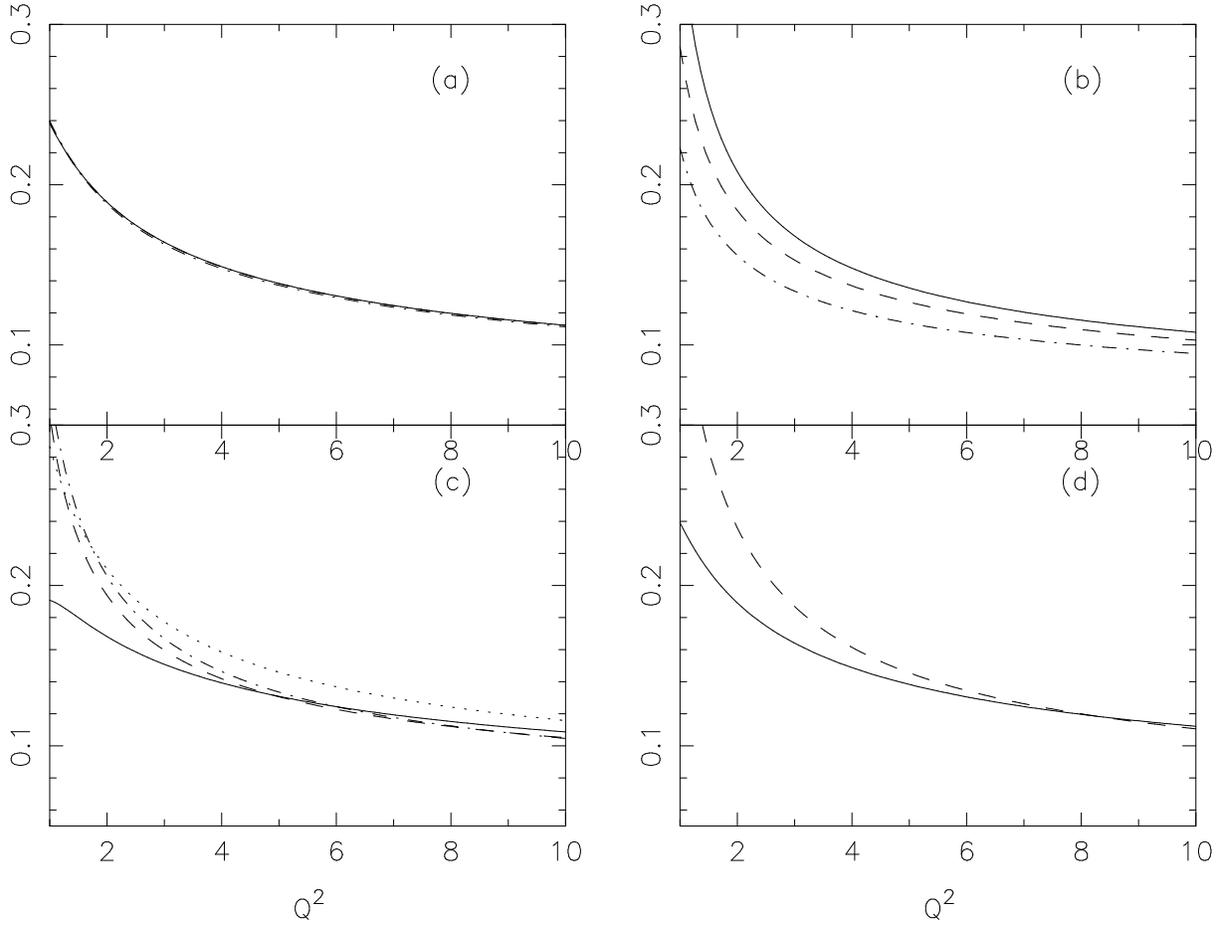}
\vspace{0.3cm}
\caption{\footnotesize Borel resummed and OPE amplitudes versus
$Q^2$ (GeV$^2$). $\alpha_s(2 {\rm GeV}^2)=0.35$ is assumed. 
(a): Borel resummed amplitudes of the perturbative part ${\rm Re}
[\Delta_{\rm P}(Q^2)]$
at NLO, NNLO, and N$^3$LO; (b): NLO (dot--dashed), NNLO (dashed), and 
N$^3$LO TPS (solid) of
$W_0(\alpha_s(Q))$;
(c): Borel resummed $\Delta_{\rm P}(Q^2)\!+\!\Delta_{\rm NP}(Q^2)$ 
(solid line)
against NLO (dashed) and NNLO (dot--dashed) OPE amplitudes. Dotted line
denotes the Borel resummed with the wrong sign;
(d): Borel resummed ${\rm Re}[\Delta_{\rm P}(Q^2)]$ 
(solid) versus an NLO OPE amplitude
(dashed).}
\label{fig-bo}
\end{figure}

We first note the remarkable stability of the Borel resummed amplitudes 
over the order of perturbation involved 
in their calculation. In Fig. \ref{fig-bo} (a)-(b), we plot, over the
interval $1\!<\!Q^2\!<\!10,$ in ${\rm GeV}^2$, the real part of the
Borel resummed
amplitudes for  $\Delta_{\rm P}(Q^2)$ using
NLO, NNLO, and N$^3$LO  perturbations  and the corresponding
amplitudes of the ordinary TPS's in the OPE approach.
In all of the plots in Fig. \ref{fig-bo}
the N$^3$LO QCD $\beta$-function was used 
in the running of the strong coupling. We also take the RG scale
at $\mu^2\!=\!Q^2$, since the RG scale dependence is sufficiently small 
for our present discussion.
The aforementioned stability of the Borel 
resummed amplitudes becomes clear when 
the two figures are compared. Note the variation
in the Borel resummed amplitudes is very small, whereas the TPS amplitudes
have significant order dependence. While this stability 
is not completely unexpected, because the leading renormalon singularity
is effectively softened by the use of the function $R(b)$ in the Borel
integration, the degree of the stability is still remarkable. This
suggests that the renormalon--induced asymptotic behavior of the 
perturbative coefficients sets in quite early in perturbation, and that
the use of $R(b)$ and conformal mapping in  Borel resummation is
very efficient in handling the renormalon singularity.
We also note in passing that the 
stability of the Borel resummed amplitudes
for  $\Delta_{\rm P}(Q^2)\!+\! \Delta_{\rm NP}(Q^2)$, which are not
shown in the figure, is comparable to 
that of $\Delta_{\rm P}(Q^2)$.

In the previous section we have seen there is a 
significant difference between the Borel resummation
and the OPE approach in the prediction of the strong coupling constant.
While there are obvious differences in the two approaches, it was not
clear what aspect of the Borel resummation is primarily responsible for 
the difference. Is it because of the perturbative part $\Delta_{\rm P}(Q^2)$
or because of our specific implementation
of the nonperturbative part $\Delta_{\rm NP}(Q^2)$, or both?

To answer this question we plot in Fig. \ref{fig-bo} (c) the Borel resummed
$\Delta_{\rm P}(Q^2)\!+\! \Delta_{\rm NP}(Q^2)$ against the OPE
amplitudes $({\rm NLO \,\, TPS}) \!+\! 0.1 /Q^2$,  
whose power term representing the
twist--four contribution is from the 
sum rule calculation \cite{braun-ko},\footnote{We note, however, there are
some variations in the estimate of the twist--four contribution.
The sum rule calculation of Ref.~\cite{Ross:1993gb} predicts 
the ($d\!=\!2$) power term $0.16 \ {\rm GeV}^2/Q^2$.}
and $({\rm NNLO \,\, TPS}) + 0.02/Q^2$. 
The small power term in the latter was
chosen for the amplitude to match the NLO OPE amplitude at large
values of $Q^2$ in the plots. In the figure we first notice that 
the NLO and  NNLO OPE
amplitudes with a large difference in twist--four 
contribution agree reasonably
well over the whole range of $Q^2$ considered. This implies that the 
higher--twist
term in the NLO amplitude can be largely accounted for by the NNLO 
perturbative term, which is in qualitative 
agreement with the observation in 
Refs.~\cite{KPS,yang}.
This also shows that the use of the NLO sum rule calculation
of the higher--twist contribution 
with a TPS of different order, which is not
an uncommon practice, can be dangerous. Higher--twist contributions 
calculated at a given order of the leading perturbative contribution 
should never be used with a TPS of different order.

On the other hand, the Borel resummed amplitude is in a 
reasonably good agreement
with the OPE amplitudes at large  $Q^2 ( > 4\, {\rm GeV}^2$), but deviates
significantly at small momenta. Obviously, this deviation at small momenta
explains the difference in the prediction of the strong coupling.
Before we answer the origin of this deviation, we note that
the good agreement of the two approaches
at large momenta is a nontrivial result. Even though the amplitudes from
the two approaches should agree 
at very high momenta (or at small couplings),
since they have the same low order perturbations,
the degree of agreement seen here is unlikely to be
a random consequence. For
instance, if our  $\Delta_{\rm NP}(Q^2)$ had the wrong sign or were 
zero, then we would see a significant difference at high momenta (see the
dotted plot for the wrong sign case).
Thus this good agreement of our Borel resummed amplitude with the
OPE amplitudes at high momenta may be regarded as a partial
support for our prescription of the nonperturbative part.

Now back to the question of what causes the deviation at low momenta.
It is not difficult to see  $\Delta_{\rm P}(Q^2)$ must be responsible
for the deviation. The reason is as follows. 
Since  the nonperturbative part $\Delta_{\rm NP}(Q^2)$ essentially behaves 
like a power suppressed term,\footnote{
$\Delta_{\rm NP}(Q^2) \sim \alpha_s(Q^2)^{\gamma_2/\beta_0}/Q^2
= \alpha_s(Q^2)^{32/81}/Q^2$, in accordance with 
Eqs.~(\ref{np})--(\ref{masterequation}) and the
OPE calculation of Ref.~\cite{Shuryak:1981kj}.} 
and $\Delta_{\rm P}(Q^2)\!+\! \Delta_{\rm NP}(Q^2)$
is in  agreement with the OPE amplitudes at 
large momenta, so should it be at low momenta, too, were 
$\Delta_{\rm P}(Q^2)$  to 
behave like an OPE amplitude. Thus the primary cause of the difference
in the predicted strong coupling constant must be that the Borel resummed 
$\Delta_{\rm P}(Q^2)$ at low momenta cannot be parametrized in the 
form of an OPE amplitude. We can see this explicitly by looking at the
plots in Fig. \ref{fig-bo} (d) where, as an example,
the NNLO ${\rm Re}[\Delta_{\rm P}(Q^2)]$ is plotted against an OPE amplitude
$({\rm NNLO \,\,TPS}) \!+\! 0.16/Q^2$. 
The power suppressed term in the latter was
fixed so that the two amplitudes 
match at high momenta in the plots. Clearly,
they deviate significantly at low momenta, with the Borel resummed growing
more slowly than the OPE amplitude as the momentum is decreased. 

\begin{figure}[ht]
 \centering\epsfig{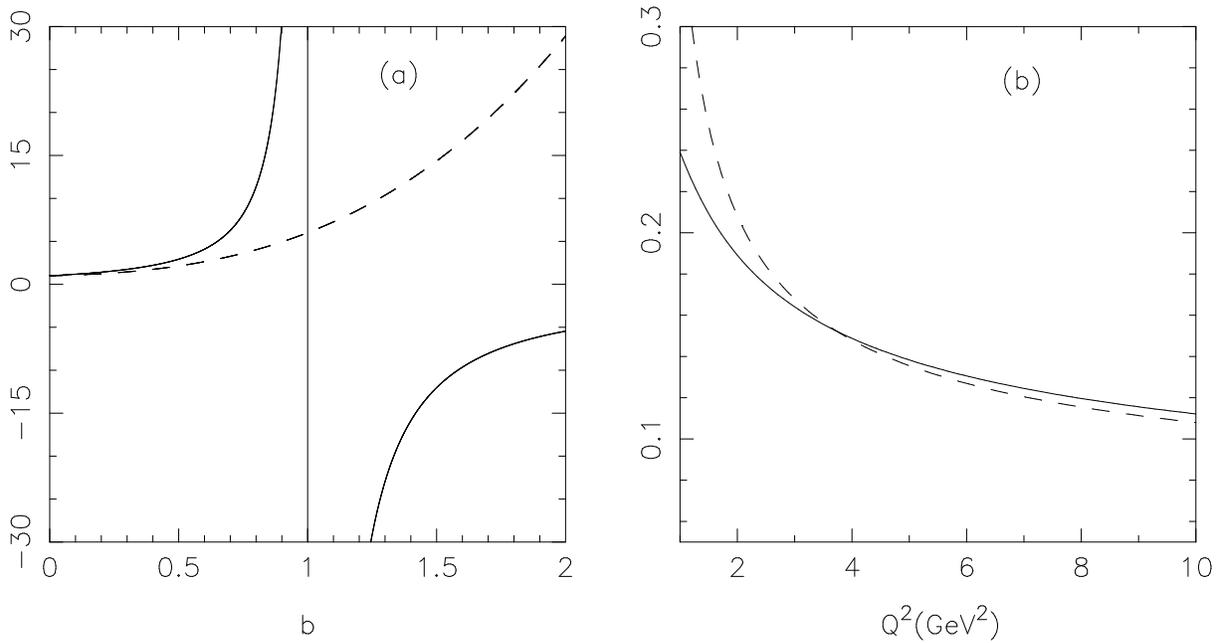}
\vspace{0.3cm}
\caption{\footnotesize (a): N$^3$LO Borel transforms. The solid line
represents the real part of the Borel transform with the renormalon 
at $b=1$ properly taken into account, and the dashed line represents 
the TPS Borel transform. (b): N$^3$LO Borel 
resummed ${\rm Re}[\Delta_{\rm P}(Q^2)]$ (solid line)
versus N$^3$LO TPS (dashed). }
\label{fig-bo1}
\end{figure}

This suggests the OPE amplitude tends to overestimate at low momenta.
We can easily see that this tendency arises from the
bad functional form of the polynomial Borel transform (TPS of
(\ref{boreltransform}))
around the renormalon singularity at $b=1$.
In Fig. \ref{fig-bo1} (a) the N$^3$LO  Borel transforms in the
two approaches are plotted in the  variable $b$.
For $b>1$ the Borel transform in our approach
defined through $R(b)$ in (\ref{rb-new})
is complex, and its real part is plotted.
It is obvious that the OPE Borel transform is badly broken around and
beyond the renormalon singularity. When the coupling is small 
this is not a serious problem because the dominant contribution to the
Borel integral (\ref{borelintegral}) comes from the region close
to the origin. However, as the coupling becomes larger the relevant
integration region extends to the renormalon singularity, and beyond,
 and as we see
in the plots, the OPE Borel transform can grossly overestimate 
at large couplings. The amplitudes obtained from these Borel transforms 
are plotted in Fig. \ref{fig-bo1} (b). As expected, at small momenta
(large coupling) the OPE amplitude is larger than
the Borel resummed. Note, on the other hand, at high momenta 
it is smaller than the latter.
This is because 
the OPE Borel transform in the region $0\!<b\!<1$, from which the
dominant contribution comes at small couplings, is smaller than
the other one, which is a characteristic feature 
rendered automatically by the correct 
implementation of the renormalon singularity in the latter.
This difference between the Borel resummed and the OPE amplitude
at small couplings may be regarded as the resummation of the
unaccounted higher order terms in the same sign asymptotic series.

That the Borel resummed $\Delta_{\rm P}(Q^2)$
cannot be approximated by an OPE amplitude of 
a TPS plus a power 
suppressed term representing the renormalon effect may appear 
contradictory to the common opinion which states otherwise. 
The latter opinion, which is based on the factorially growing large order
behavior and the running coupling, would be true in a sense,
provided the strong coupling were sufficiently small, and a TPS of
large order ($\sim 1/\alpha_s$) was used. In reality, however,
the strong couplings at the low momenta we consider are not so small, and
there is no guarantee that the Borel resummed $\Delta_{\rm P}(Q^2)$
at those momenta can
be parameterized as an OPE amplitude.
The example here clearly shows that
cannot be generally true. A Borel resummed amplitude can have a much
more complex functional behavior than the sum of a TPS and a power 
term intended to account for the renormalon effect.
This consideration suggests that 
several existing analyses of low energy QCD 
observables based on the OPE approach should be reexamined, since the 
issues raised here are likely to be relevant there, too.

To sum up this section, 
we have made two observations concerning the Borel resummation
and the OPE approach.
First,  our method of calculation of the perturbative plus nonperturbative
contribution in Borel resummation  is
consistent at larger $Q^2 > 4 \ {\rm GeV}^2$
with the OPE approach using QCD sum rule calculation, 
and secondly, 
at low energies the OPE amplitude tends to overestimate, and
the Borel resummed amplitude with a proper incorporation of the
leading renormalon {\it cannot} 
be approximated by an OPE amplitude of a TPS 
with a power suppressed term. It is the second observation that
directly accounts for the differences in the extracted strong coupling
constants from the two approaches.

\section{Conclusions}
\label{conclusions}

We performed a resummation of the Gross--Llewellyn Smith (GLS) 
sum rule by fully accounting for the correct known form of the leading
infrared (IR) renormalon singularity at $b\!=\!1$ in the Borel plane.
As one direct consequence of this singularity,
the resummed ``perturbative'' part 
$\Delta_{\rm P}(\alpha_s(Q))$ of the GLS has a branch cut of the
form $(- \alpha_s(Q^2))^{-\nu} \exp(- \pi/\beta_0 \alpha_s(Q))
[ 1 + {\cal O}(\alpha_s) ]$,
i.e., a twist--four term 
$(-1)^{- \nu} (1/Q^2) \alpha_s^{\gamma_2/\beta_0} [ 1+{\cal O}(\alpha_s) ]$
with the branch cut discontinuity factor $\exp ( \pm i \pi \nu)$.
Here, $\gamma_2\!=\!8/9$ is the known one--loop coefficient of the
anomalous dimension of the corresponding twist--four $d\!=\!2$
operator appearing in the OPE.
A crucial element of the analysis was the fixing of the
``nonperturbative'' part as the negative of the 
aforementioned branch cut term, thus making the
resummed (``perturbative'' plus ``nonperturbative'')
GLS sum rule manifestly real. This procedure is free from
the known ambiguity of separation of the ``perturbative''
and ``nonperturbative'' parts.
In the Borel resummation of the ``perturbative''
part, we further employed a conformal transformation
to minimize the numerical influence of other
renormalon singularities.
All this allowed us to perform the resummation of the
massless part of the GLS sum rule, i.e., of the
contributions of the three light quark flavors.
The contributions of the heavy ($c$--)quark
were added separately, as were the target nuclear correction
contributions and the light--by--light contributions. 
These three types of contributions turned out to be small,
in contrast to the ``nonperturbative'' contributions
which turned out to be significant.
The calculations were performed in the
${\overline {\rm MS}}$ renormalization scheme,
and the renormalization scale $\mu^2$ ($\sim Q^2$) was taken
in the region of the smallest $\mu^2$--sensitivity of our results.

We then confronted the resummed expressions with the 
Fermilab CCFR Collaboration data \cite{ccfr} 
for the GLS (at $Q^2 = 2, 3.16 \ {\rm GeV}^2$)
which already include the target mass
corrections. Our central value prediction for the
QCD coupling parameter, corresponding to the central
GLS values of the CCFR, is $\alpha_s(M_Z) \approx 0.117$
[see Eqs.~(\ref{al2o2}) and (\ref{al3o3})],
in good agreement with the present world average. 
This is different from the central value
predictions of previous analyses of the GLS
sum rule by the CCFR Collaboration \cite{ccfr} 
($\alpha_s(M_Z) \approx 0.114$) and 
by the authors of \cite{Chyla:1992cg}
($\alpha_s(M_Z) \approx 0.115$) which
are below or at the lower edge of the world average.
We have seen that our approach to the calculation
of the nonperturbative contribution is consistent with
the OPE approach, and the main reason for the difference
between the two approaches is that at small $Q^2 < 4 \ {\rm GeV}^2$,
the OPE approach tends to overestimate and
the Borel resummed perturbative contribution cannot
be approximated by an OPE amplitude.

The GLS sum rule, at the low gauge boson transfer
momenta $Q^2\!=\!2$--$4 \ {\rm GeV}^2$, is a very
important quantity to measure, because it has
apparently a strong nonperturbative component,
stronger than in some other low--energy QCD observables
such as the semihadronic tau decay rate.
The more precise experimental values of the
GLS sum rule would help determining 
the higher--twist contributions more accurately.

\acknowledgments
The work of C.C. was supported by DGIP (UTFSM).
The work of G.C. was supported by the FONDECYT (Chile)
Grant No.~1010094. The work of
T.L. and K.S.J. was supported in part by the BK21 Core Program.

\begin{appendix}

\section[]{Branch cut singularity and the nonperturbative part}
\setcounter{equation}{0}
\label{appendix2}

In this Appendix, we show explicitly formulas leading to 
Eqs.~(\ref{branchcut})--(\ref{masterequation}), using
the identity (\ref{borelintegral}) and the leading IR
renormalon singularity structure (\ref{singularity}).
The latter structure around $b\!=\!1$ can be rewritten 
more explicitly as
\begin{eqnarray}
{\tilde W}_0(b) &=& \frac{C}{\Gamma(-\nu)} \beta_0^{1+\nu}
(1 - b)^{-1 - \nu} \left[ 1 + \kappa_1 (1 - b) + \kappa_2 (1 - b)^2
+ \cdots \right] 
\nonumber\\
&& + ({\rm Analytic} \ {\rm part})  \ .
\label{singularity2}
\end{eqnarray}
The leading part is known ($\nu$), while the
coefficients $\kappa_j$ ($j\geq1$) of the subleading
parts are not yet known. 
Inserting the above expansion into the Borel
integration formula (\ref{borelintegral}) and performing
the change of the integration variable $b = 1 + \beta_0 a t$,
we obtain
\begin{equation}
{\rm Im} \Delta_{\rm P}(a(Q) \pm {\rm i} \epsilon) =
\frac{C}{\Gamma(-\nu)} e^{-1/\beta_0 a(Q)} a(Q)^{- \nu}
f_{\pm}(a(Q)) \ ,
\label{DP1}
\end{equation}
where 
\bear
f_{\pm}(a) &=& {\rm Im}\int_{0 \pm {\rm i} \epsilon}^
{\infty \pm {\rm i} \epsilon} dt \; e^{-t} (-t)^{-1-\nu}
\left[ 1 + \kappa_1 (\beta_0 a) (-t) + \kappa_2 (\beta_0 a)^2 (-t)^2 +
\cdots \right] \nonumber \\
&=&\mp \sin(\pi\nu)\int_{0 }^
{\infty } dt \; e^{-t} t^{-1-\nu}
\left[ 1 + \kappa_1 (\beta_0 a) (-t) + \kappa_2 (\beta_0 a)^2 (-t)^2 +
\cdots \right] \nonumber \\
&=&\mp\Gamma(-\nu) \sin( \pi \nu) \left[ 1 + \kappa_1  (\beta_0 a) \nu
+ \kappa_2 (\beta_0 a)^2 \nu (\nu\!-\!1) + \cdots \right]
\label{fpm} 
\eear
Now  requiring the imaginary parts 
of $\Delta_{\rm P}$ at positive $a(Q)$ to match those in (\ref{DP1}) 
$\Delta_{\rm P}$ for general complex $a(Q)$ can be written as
\begin{eqnarray}
\Delta_{\rm P}(a(Q)) &=&
- C e^{-1/\beta_0 a(Q)}  \Big\{ ( - a(Q))^{- \nu}
+ \kappa_1 (- \nu) \beta_0 (-a(Q))^{-\nu +1} 
\nonumber\\
&&+ \kappa_2 (- \nu) (-\nu\!+\!1) \beta_0^2 (-a(Q))^{-\nu+2} + \cdots
\Big\}
\nonumber\\
&&+ {\tilde {\Delta}}_{\rm P}(a(Q)) \ ,
\label{DP2}
\end{eqnarray}
where ${\tilde {\Delta}}_{\rm P}(a)$ is the part with no
singularities (no cuts) for $a > 0$. \footnote{
The absolute value of the singular term in (\ref{DP2})
can be rewritten as $\sim (\Lambda^2/Q^2) a(Q)^{+\gamma_2/\beta_0} [ 1 +
{\cal O}(a) ]$, where $\Lambda$ is a $Q$--independent scale,
and $\gamma_2$ is the one--loop coefficient of the
anomalous dimension of the twist--four ($d\!=\!2$)
operator $\ll O(Q) \gg \sim \Lambda^2 a(Q)^{+\gamma_2/\beta_0}$
appearing in the OPE (\ref{ope}).}
According to Ref.~\cite{lee5}, the nonperturbative
part $\Delta_{\rm NP}(a(Q) \pm {\rm i} \epsilon)$ must cancel the
imaginary part of $\Delta_{\rm P}(a(Q) \pm {\rm i} \epsilon)$,
and $\Delta_{\rm NP}$  was chosen to have the simplest,
presumably the most natural, form -- i.e.,
just the negative of the branch cut term of (\ref{DP2})
\begin{eqnarray}
\Delta_{\rm NP}(a(Q)) &=&
+ C e^{-1/\beta_0 a(Q)}  \Big\{ ( - a(Q))^{- \nu}
+ \kappa_1 (- \nu) \beta_0 (-a(Q))^{-\nu +1} 
\nonumber\\
&&+
\kappa_2 (- \nu) (-\nu\!+\!1) \beta_0^2 (-a(Q))^{-\nu+2} + \cdots
\Big\} \ .
\label{DNP2}
\end{eqnarray}
Further, 
since $(-a \mp {\rm i} \epsilon)^{- \nu + n} =
a^{-\nu + n} \exp[ \pm {\rm i} (\nu\!-\!n) \pi]$,
expressions
(\ref{DP1}) and (\ref{DNP2}) immediately relate
${\rm Re} \Delta_{\rm NP}(a\pm {\rm i} \epsilon)$
with the (calculable) ${\rm Im} \Delta_{\rm P}(a\pm {\rm i} \epsilon)$
\begin{equation}
{\rm Re} \Delta_{\rm NP}(a\pm {\rm i} \epsilon) = \mp \cot (\nu \pi) 
{\rm Im} \Delta_{\rm P}(a\pm {\rm i} \epsilon) \ ,
\label{NPvsP}
\end{equation}
giving the result (\ref{masterequation}). 

\end{appendix}

\noindent

\newpage

\noindent

\noindent

\noindent

\noindent

\noindent

\noindent

\noindent

\end{document}